\documentclass[a4paper]{article}

\usepackage{moreverb}

\usepackage[dvips,colorlinks,bookmarksopen,bookmarksnumbered,citecolor=red,urlcolor=red]{hyperref}

\newcommand\BibTeX{{\rmfamily B\kern-.05em \textsc{i\kern-.025em b}\kern-.08em
T\kern-.1667em\lower.7ex\hbox{E}\kern-.125emX}}

\usepackage{times,a4wide}
\usepackage{tikz}
\usepackage{verbatim}
\usetikzlibrary{%
    decorations.pathreplacing,%
    decorations.pathmorphing%
}

\usepackage{tikz}
\usetikzlibrary{calc,patterns,decorations.pathmorphing,decorations.markings}
\usepackage{verbatim}
\usepackage{tikz}
\usetikzlibrary{%
    decorations.pathreplacing,%
    decorations.pathmorphing%
}

\usetikzlibrary{calc,patterns,
                 decorations.pathmorphing,
                 decorations.markings}

\usetikzlibrary{decorations.pathmorphing} % for snake lines
\usetikzlibrary{matrix} % for block alignment
\usetikzlibrary{arrows} % for arrow heads
\usetikzlibrary{calc} % for manimulation of coordinates

\tikzstyle{block} = [draw,rectangle,thick,minimum height=2em,minimum width=2em]
\tikzstyle{sum} = [draw,circle,inner sep=0mm,minimum size=2mm]
\tikzstyle{connector} = [->,thick]
\tikzstyle{line} = [thick]
\tikzstyle{branch} = [circle,inner sep=0pt,minimum size=1mm,fill=black,draw=black]
\tikzstyle{guide} = []
\tikzstyle{snakeline} = [connector, decorate, decoration={pre length=0.2cm,
                         post length=0.2cm, snake, amplitude=.4mm,
                         segment length=2mm},thick, magenta, ->]

\usepackage{mathrsfs}

\usepackage{amsmath}
\usepackage{mathtools}

\usepackage[normalem]{ulem}

\usepackage{amsmath,amssymb}
\usepackage{graphicx}
\usepackage{color}
\usepackage{indentfirst} %%%Ê׶οոñ

\usepackage{booktabs}%%%excel biaoge ºê
\usepackage{multirow}

\usepackage{subfigure}

\newtheorem{definition}{Definition}
\newtheorem{assumption}{Assumption}
\newtheorem{theorem}{Theorem}
\newtheorem{remark}{Remark}

\newtheorem{lemma}{Lemma}
\newtheorem{corollary}{Corollary}

\newtheorem{problem}{Problem}

\begin{document}

%\runningheads{A.~N.~Other}{A demonstration of the \journalabb\
%class file}

\title{Gain-scheduled Leader-follower Tracking Control for Interconnected
  Parameter Varying Systems\thanks{Accepted for publication in
    International Journal of Robust and Nonlinear Control} \thanks{This work was supported by the Australian Research Council under Discovery Projects funding scheme (projects DP0987369 and DP120102152).}}

\author{Yi Cheng and V. Ugrinovskii\thanks{Corresponding author.}}
\date{\normalsize \em School of Engineering and Information Technology,\\The University of New South Wales at
  the Australian Defence Force Academy, Canberra, Australia,\\
  Email:\{yi.cheng985,v.ugrinovskii\}@gmail.com} 

\maketitle

\begin{abstract}
This paper considers the gain-scheduled leader-follower tracking control problem for
a parameter varying complex interconnected system with directed communication topology
and uncertain norm-bounded coupling between the agents. A gain-scheduled
consensus-type control protocol is proposed and a sufficient condition is
obtained which guarantees a suboptimal bound on the system
tracking performance under this protocol. An interpolation technique is used to
obtain a protocol schedule which is continuous in the scheduling
parameter. The effectiveness of the proposed method is demonstrated using a
simulation example.
 \vspace{1cm}

\end{abstract}

\section{Introduction}

In recent years, the topic of cooperative control has attracted
much attention. The objective of the cooperative control problem is to propose
distributed control laws to achieve a desired system behavior \cite{Shamma2007}. A related problem is that of synchronization of complex dynamical systems
where all the components are controlled to exhibit a
similar behavior by interconnecting them into a network
\cite{[Tuna2008],[Tuna2009]}.

There are several approaches to the synchronization problem for
complex systems consisting of many dynamic subsystems-agents. In the average consensus problem,
which has received considerable attention in the last decade
\cite{Olfati2004, Olfati2007}, the objective is to synchronize all the
agents to a common state. Another approach is to employ a suitable internal model for the system. For instance, it is shown in
\cite{Wieland2011}  that the
existence of such an implicit internal
model is necessary and sufficient for synchronization of a system of
heterogeneous linear agents considered in that paper.
However, the internal model approach does not directly address the overall
system performance, since each system is controlled to follow its own
internal model dynamics. In general, it may be difficult to determine an
internal model that guarantees a good synchronization performance. Yet another
approach is to designate one of the agents to serve as a leader, and design
interconnections within the system so that the rest of the system follows
the leader~\cite{Pecora1990}; also, see \cite{Grip2012} for a recent
example. This idea leads to the \emph{leader-follower tracking}
problem, which is the main focus of this paper.

A common feature of many papers that consider the leader-follower problem
is that the dynamics of agents are usually assumed to be dynamically
decoupled~\cite{Scardovia2009,Jadbabaie2003,Ren2007,Li2010,Hong2006,Hu2007}. In many complex systems, however, interactions between
subsystems are inevitable and must be taken into account
\cite{Siljak1991}. Examples of
systems with dynamical interactions between subsystems include spacecraft
control systems and power systems~\cite{Siljak1978}. For
example, it was noted in \cite{Pota2006} that
weak uncertain couplings between generators is one of the reasons for
dynamic instability in power systems and, therefore, they should not be
neglected. This motivates us to consider the leader-follower
tracking problem for interconnected systems. While we do not
consider a specific application, our approach makes a step towards
using consensus feedback for synchronization of such systems, compared to
other contributions in the area of networked control systems which do not
address the presence of interconnections.

In this paper, we are concerned with the leader-follower tracking problem for interconnected systems that depend on a time varying parameter. Analysis and control of linear parameter varying systems has attracted much attention in the last two decades due to their
applications in flight control \cite{Shamma2012}, turbofan engines and wind
turbine systems \cite{Bianchi2006,MSK-2012}.
For example, an application of the distributed control approach to
control and synchronization of wind generation systems modeled as parameter varying systems has been presented in~\cite{MSK-2012}.
Even though a discrete time model was considered in \cite{MSK-2012}, parameter varying system modeling was motivated by the fact
that the wind energy source driving wind
turbines exhibits time-varying nature. In order to integrate a wind
generation system into a power grid in a grid-friendly manner, the total
power output of the wind turbines must be regulated to conform to a
constant output. Each turbine then serves as a node and is controlled with
respect to its power output by turning the blade pitch angle, and the value
of the blade pitch angle is propagated through the communication network
using the leader-follower algorithm. Thanks to this and many other
potential applications, the theory of cooperative control for parameter
varying systems has been gaining attention in recent years
\cite{Seyboth2012a,Liu2013}.

The main contribution of this paper concerns the leader-follower control of
parameter varying interconnected systems. A related problem, from the
synchronization viewpoint, has been studied in \cite{Seyboth2012,U8}. In
the first reference, the synchronization problem is solved for
heterogeneous systems which depend on the parameters in an affine fashion,
under the assumption that
the network of LPV agents allows for an internal model for synchronization
while the agents are decoupled from each other. That is, the agents
interact over the control protocol only.
In~\cite{U8},
while the linear dependency on the parameter is not required, the
leader is assumed to be given and be completely decoupled from the
agents. In contrast with these references, we assume the leader to be
chosen from the group of parameter varying agents; it
is interconnected to the rest of the network, and the linear dependency on
the parameter is not required.

The fact that the leader is interconnected with the followers makes it
difficult to apply regular centralized tracking techniques to the problem
under consideration. Many tracking techniques assume that the reference
trajectory is generated by an exosystem  or is \emph{a priori} known, and
is independent of the followers. This is not the case in this
paper. Alternatively, a centralized tracking controller can be
obtained by solving a stabilization problem for a large-scale system
comprised of subsystems describing dynamics of individual tracking
errors. This can be done by applying one of the existing centralized
gain-scheduling robust control techniques, e.g., using the technique from
\cite{[Yoon2007]}. However, in general the centralized controller obtained
this way will not have a desired information structure. Indeed, a controller
obtained in such a way will not generally guarantee that the information
from subsystems that are not observed by a node is not required for
feedback. One way to enforce such an information constraint is to employ a
block diagonal Lyapunov function. This is the solution approach undertaken
in this paper and is its main difference from solutions which could be
obtained in a centralized setup.

Different from many papers that study
synchronization of decoupled systems (including the above mentioned
references~\cite{Seyboth2012,U8}, also see \cite{[Tuna2008],[Tuna2009]}), in
the case of coupled systems, it is essential to distinguish the network
representing the existing interactions between the subsystems
(including  interactions with the leader) from the network which
realizes communication and control.
This leads us to consider a two-network structure in this paper. The
rationale for this is twofold. Firstly, our aim is to construct synchronization protocols for all
agents excluding the leader, as we wish to avoid perturbing the
leader dynamics (other than through an unavoidable physical
coupling). Hence, the communication graph of the network must be different from the
interconnection graph. The second reason is that the interconnection graph
describes dynamical couplings between subsystems and thus may be different from
the communication and control graph.

Our main result is a sufficient condition
for the design of a gain-scheduled leader-follower control protocol for
parameter varying multi-agent systems with directed control network
topology and linear uncertain couplings subject to norm-bounded
constraints. The condition involves checking feasibility of
parameterized linear matrix inequalities (LMIs) at several operating points
of the system. These LMIs serve as the basis
for the design of a continuous (in the
scheduling parameter) control protocol for
 coupled parameter varying systems, by interpolating consensus control protocols
 computed for those operating points;
 cf.~\cite{Stilwell1999,U8,[Yoon2007]}. Interpolation
 allows us to mitigate
 detrimental effects of transients arising when the system traverses from
 one operating condition to another. % A Lyapunov function constructed for

The remainder of the paper proceeds as follows. In Section \ref{problem formulation}, we formulate the leader
follower control problem for parameter varying coupled multi-agent systems
and give some preliminaries. The main results are given in Section \ref{main}. Section \ref{example} gives an example
which illustrates the theory presented in the paper. Finally, the
conclusions are given in Section \ref{conclusions}.

\section{Problem Formulation and Preliminaries}
\label{problem formulation}

\subsection{Interconnection and communication graphs}

\begin{figure}
 \centering
 \scalebox{0.7}{
 \begin{tikzpicture}[->,>=stealth',shorten >=1pt,auto,node distance=2cm,
   %thick,main node/.style={circle,fill=blue!20,draw,font=\sffamily\Large\bfseries}]
   thick,main node/.style={circle,fill=white!20,draw,font=\sffamily\bfseries},]

   \node[main node] (1) {1};
   \node[main node] (2) [right of=1] {2};
   \node[main node] (3) [right of=2] {3};
   \node[main node] (4) [below of=2] {4};
   \node[main node] (5) [above of=2]{0};

\path[every node/.style={font=\sffamily\small}]

    (5) edge [bend right] node {$d_1$} (1)
    (1) edge [bend right]  (2)
    (2) edge [bend right]  (3)
    (3) edge [bend left]  (4)
    (3) edge [bend right] node {$\bar d_3$} (5)
    (3) edge [bend right]  (2);

   \path[draw,blue,dashed] (5)  -- node {$g_2$}(2);
   \path[draw,blue,dashed] (2)  -- (1);
   \path[draw,blue,dashed] (2)  -- (3);

  \draw[thick, blue, dashed]
    (2.1,-0.32)-- (2.1,-1.66);

     \draw[blue, dashed]
    (1.9,-1.63)-- (1.9,-0.3);

  \draw[blue, dashed]
    (5.2,0)-- node[above =0.08cm, text width=5cm,text centered]
     {\textcolor{black}{Communication edges}}(7.5,0);

  \draw[black]
  (5.2,1.2)-- node[above =0.08cm, text width=5cm,text centered]
     {Interaction edges}(7.5,1.2);
 \end{tikzpicture}
 }
 \centering
 \caption{An example of the interconnection and communication graphs.}
 \label{example graph}
 \end{figure}
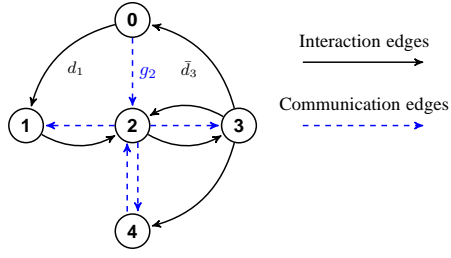

As stated in the introduction, in this paper we draw a distinction between
the communication topology of the system used for control and the topology
of interactions between subsystems. The example of the two-graph structure
is shown in Fig.~\ref{example graph}, where the edges of the
interconnection graph are indicated by the solid lines and the edges
of the communication graph are shown by the dashed lines. Both coupling and communication topologies are described in terms of directed graphs defined on the common node set
$\mathcal {V}= \{0, \ldots, N\}$. Without loss of generality, node 0
will be assigned to be the leader of
the network, while the nodes from the set $\mathcal {V}_0=\{1, \ldots, N\}$
will represent the followers.
The coupling graph and the communication graph will be denoted as
$\mathcal {G}^\varphi$ and
$\mathcal {G}^c$,
respectively.

The edge sets of both graphs are subsets of the set $\mathcal{V}\times \mathcal
{V}$ and consist of pairs of nodes. The pair $(j,i)$ in
each edge set denotes the directed edge which originates at node $j$ and
ends at node $i$. Edge $(j,i)$ in the edge set of the
directed coupling graph $\mathcal {G}^\varphi$, denoted
$\mathcal {E}^\varphi$, describes the fact that node $i$ is influenced by node
$j$ through a directed interaction between nodes $j$ and $i$.

Also, $\mathcal {E}^c \subseteq \mathcal {V}\times
\mathcal {V}$ denotes an edge set of the communication graph
$\mathcal {G}^c$, consisting of ordered pairs of nodes. Each such edge indicates the information flow between the
nodes, that is, $(j,i)\in \mathcal{E}^c$ if and only if node $i$ obtains
information from node $j$, which it can use for control.

The adjacency matrix of the directed interconnection graph
$\mathcal {G}^\varphi$ is denoted as $\mathcal{A}^\varphi$, its
$(i, j)$-th entry is 1 if and only if $(j, i)\in \mathcal{E}^\varphi$. The adjacency matrix
$\mathcal{A}^c$ of the directed communication graph $\mathcal {G}^c$ is defined
in the same manner. Since
according to a standard convention we assume that both the coupling graph
and communication graph have no self-loops, the diagonal entries of
$\mathcal{A}^\varphi$ and $\mathcal{A}^c$  are all equal to zero.

To distinguish between the leader and the rest of the
network, we define subgraphs $\mathcal {G}^\varphi_0$, $\mathcal {G}^c_0$  of the graphs
$\mathcal {G}^\varphi$, $\mathcal {G}^c$ defined on the node set $\mathcal {V}_0=\{1, \ldots,
N\}$, with the edge sets $\mathcal {E}_0^\varphi=\{(j,i)\in \mathcal
{E}^\varphi\colon i,j\in \mathcal {V}_0\}$ and $\mathcal {E}_0^c=\{(j,i)\in
\mathcal{E}^c\colon i,j\in \mathcal {V}_0\}$, respectively. Selecting the subgraphs $\mathcal {G}^\varphi_0$  and $\mathcal {G}^c_0$ induces
the partition of the matrices $\mathcal{A}^\varphi$, $\mathcal{A}^c$,
\begin{eqnarray*}
&&\mathcal{A}^\varphi
=\left[\begin{array}{c|c}
0 & \bar d \\
\hline
d & \mathcal{A}^\varphi_0
\end{array}
\right], \quad \mathcal{A}^c
=\left[\begin{array}{c|c}
0 & 0 \\
\hline
g & \mathcal{A}^c_0
\end{array}
\right],
\end{eqnarray*}
where $d=[d_1~ \ldots~ d_N]'$, $\bar d=[\bar d_1~
\ldots~ \bar d_N]$, $g=[g_1~ \ldots~ g_N]'$ and
$\mathcal{A}^\varphi_0$, $\mathcal{A}^c_0$ are the adjacency
matrices of the subgraphs $\mathcal {G}^\varphi_0$ and $\mathcal {G}^c_0$,
respectively. The zero row of the matrix $\mathcal{A}^c$ reflects our
assumption that the leader node does not receive the state information
from other nodes of the network. However, the corresponding row of
$\mathcal{A}^\varphi$ may be nonzero since the leader node may be physically
coupled with some of the followers.

In this paper we are concerned with the case where only some of the
followers receive the state information directly from the leader. We refer
to such nodes $i\in\mathcal{V}_0$ as
pinned nodes; the corresponding entries of the adjacency matrix
$\mathcal{A}^c$, $g_i=1$, and $g_i=0$ if node $i$ is not pinned. The matrix
$G=\mathrm{diag}\{g_i\}\in \Re^{N\times N}$ is referred to as a pinning matrix.

The Laplacian matrix of the subgraph $\mathcal {G}^c_0$ is defined as $\mathcal {L}^c_0 =\mathcal {P} - \mathcal {A}^c_0$,
where $\mathcal {P} =\mathrm{diag}\{p_1,\ldots,p_N\}\in R^{N\times N}$ is
the in-degree matrix of $\mathcal {G}^c_0$, i.e., the diagonal matrix,
whose diagonal elements are the
in-degrees of the corresponding nodes of the graph $\mathcal {G}^c_0$,  $p_i =
\sum_{j=1}^{N}a_{ij}^c$ for $i = 1,\ldots,N$, where $a_{ij}^c$ are the
elements of the $i$th row of the matrix $\mathcal{A}^c_0$.

Finally, we give the definition and the notation for neighborhoods in the above
graphs.
\begin{definition}
 $\mathcal{G}^c_0=(\mathcal {V}, \mathcal {E}^c)$ is a communication graph, each node $i \in \mathcal {V}$ represents a subsystem in the interconnected system and edge $(j, i) \in \mathcal {E}^c$  means that node $i$ obtain information from node $j$.
\end{definition}

\begin{definition}
$\mathcal {G}^\phi$  is an interconnection graph, each node $i \in \mathcal {V}$ represents a subsystem in the interconnected system  and edge $(j, i) \in \mathcal {E}^\phi$  indicates that subsystem $i$ is influenced by subsystem $j$ through a directed interaction between subsystems $i$ and $j$.
\end{definition}

Node $j$ is called a neighbor of node $i$ in the graph
$\mathcal {G}^\varphi$ (or $\mathcal {G}^\varphi_0$, $\mathcal {G}^c$,
$\mathcal {G}^c_0$, respectively) if
$(j,i)\in \mathcal {E}^\varphi$ ($\mathcal {E}_0^\varphi, \mathcal {E}^c$ or
$\mathcal {E}_0^c$, respectively). The sets of neighbors of node $i$ in the graphs $\mathcal {G}^\varphi$ and $\mathcal {G}^c$ are denoted as $N^\varphi_i=\{j|(j,i)\in \mathcal {E}^\varphi\}$,
and $N^c_i=\{j|(j,i)\in \mathcal {E}^c\}$,
respectively. The sets of neighbors of node $i$ in the subgraphs $\mathcal {G}^\varphi_0$ and $\mathcal {G}^c_0$ are denoted as
$S^\varphi_i=\{j|(j,i)\in \mathcal {E}_0^\varphi\}$ and $S^c_i=\{j|(j,i)\in \mathcal {E}_0^c\}$,
respectively.

We conclude this subsection by specifying our standing requirements on the
communication topology of the network which are assumed to hold throughout
the paper.

\begin{assumption}
\label{graph assumption}
 The communication subgraph $\mathcal{G}_0^c$ contains a spanning tree,
 whose root node $i_r$ is the pinned node, i.e., $g_{i_r}> 0$.
\end{assumption}

\begin{remark}
Assumption~\ref{graph assumption} ensures that the information flows from the
leader to its pinned followers, as well as between other followers
\cite{Li2010,Zhang2012}.
\end{remark}

\subsection{Notation}
Throughout the paper, the following notation will be used.

$\Re^n$ and $\Re^{n\times m}$ are a real Euclidean $n$-dimensional vector
space and a space of real $n\times m$ matrices.

$\Gamma $ denotes an
interval $[\rho_{\min}, \rho_{\max}] \subset \Re$. The symbols $\rho$, $\rho_s$,
$\bar\rho_s$, $\underline{\rho}_s$, etc., will represent various points
within the interval $\Gamma$, and $\Gamma_\ell$, $\Gamma_0$ will denote
sets of points within $\Gamma$ which will be defined later in the paper.
Also, $\rho(t)$ is a scalar function, defined on $[0,\infty)$ and taking
values in the interval $\Gamma$. This function will describe a time-varying
scheduling parameter of the multi-agent system under consideration.

Unless stated otherwise, the notations $A(\rho)$, $Y_\rho$, etc., will
refer to matrices of appropriate dimension parameterized by $\rho\in\Gamma$.

For $q\in \Re^n$, $\mathrm{diag}\{q\}$ denotes the diagonal matrix with the
entries of $q$ as its diagonal elements.

$\otimes$ denotes the Kronnecker product of two matrices.

$\lambda_{\max}(\cdot)$ and $\lambda_{\min}(\cdot)$ will denote the largest
and the smallest eigenvalues of a real symmetric matrix.

$I_N$ is the $N\times N$ identity matrix, and $\mathbf{1}_N\in \Re ^N$
is the vector whose all entries are equal to $1$. When the dimension is
clear from
the context, the subscript $N$ will be suppressed, and $I$ will represent the
identity matrix of an appropriate dimension.

For two symmetric matrices $X$ and $Y$ of the same dimensions, $X\ge Y$,
($X>Y$) if and only if $X-Y$ is positive semidefinite (positive definite).

Consider the Laplacian matrix of the subgraph $\mathcal {G}^c_0$,
$\mathcal {L}^c_0$. According to~\cite{Zhang2012}, under
Assumption~\ref{graph assumption}, the matrix $\mathcal{L}^c_0+G$ is
nonsingular (also see \cite{Hu2007}), and all the entries of the vector $\vartheta=[\vartheta_1, \ldots, \vartheta_N]'= (\mathcal{L}^c_0+G)^{-1} \textbf{1}_N$ are positive. Also,
the matrix $\Theta^{-1}(\mathcal{L}^c_0+G)+(\mathcal{L}^c_0+G)'\Theta^{-1}$
is positive definite, where $\Theta\triangleq
\mathrm{diag}\{\vartheta\}$. The following
two constants will be used in the proofs of our results:
\begin{eqnarray}
\sigma&=&\frac{1}{2}\lambda_{\min}\left(\Theta^{-1}(\mathcal{L}^c_0+G)+(\mathcal{L}^c_0+G)'\Theta^{-1}
\right), \nonumber \\
\hat \lambda&=& \lambda_{\max}
\left((\mathcal{L}^c_0+G)'\Theta^{-2}(\mathcal{L}^c_0+G) \right).
\label{sigma.lambdahat}
\end{eqnarray}

\subsection{Problem Formulation}

 The system under consideration consists of $N+1$ parameter-varying
 dynamical agents,
 coupled with their neighbors; the topology of interconnections is captured
 by the directed graph $\mathcal {G}^\varphi$. Dynamics of the $i$th agent
 are described by the linear equation
\begin{align} \label{all agents dymamic}
\dot{x}_i=A(\rho(t))x_i+B_1u_i+ B_2\sum\limits_{j\in N^\varphi_i} \varphi_{ij} (t,
x_j-x_i), \quad i=0, \ldots, N,
\end{align}
where $x_i\in \Re^n$ is the state of agent $i$, $u_i\in \Re^p$ is the
control input and $\rho(t)$ is the time-varying parameter, which is
available to all agents. It is assumed that $\rho\colon [0, \infty) \rightarrow
\Gamma \subset \Re$ is a continuous
function. Also in equation~(\ref{all agents dymamic}), $B_1$ and $B_2$ are
real matrices of appropriate dimensions, and $A(\rho(t))$ is the
composition of a continuous matrix-valued function $A(\cdot)\colon
\Gamma\to \Re^{n\times n}$ and the function $\rho(t)$ defined above.

The functions $\varphi_{ij}\colon [0,\infty)\times \Re^n \to \Re^m$,
\[
\varphi_{ij}(t,x)= \Delta_{ij}(t)C_{ij}x, \quad x\in \Re^n,
\]
describe how agent $j$ influences the dynamics of agent $i$. Here
$C_{ij}\in \Re^{q_{ij} \times n}$ and $\Delta_{ij}\in
\Re^{m\times q_{ij}}$ are respectively constant and time-varying matrices.
According to the above model, we focus on linear time-varying interactions
between the
subsystems whose strengths depend on the relative state of subsystem $i$ with respect
to subsystem $j$, $x_j-x_i$.
Also, we will assume that the coefficients $\Delta_{ij}(t)$ are
uncertain, and satisfy the constraint
\begin{equation}
\label{nb}
\Delta_{ij}'(t)\Delta_{ij}(t) \leq I, \quad \forall t \in  [0, \infty).
\end{equation}
That is, we consider a class of norm-bounded uncertain
interactions, which will be denoted by $\Xi$.

Since we have designated agent 0 to be the leader, and the rest of the agents
to be controlled are to follow the leader, then according to (\ref{all agents
  dymamic}) dynamics of the $i$th follower are described by the equation
\begin{align} \label{agents dymamic}
\dot{x}_i=&A(\rho(t))x_i+B_1u_i+ B_2\Big(\sum\limits_{j\in S^\varphi_i} \varphi_{ij} (t,
x_j-x_i) + d_i\varphi_{i0} (t, x_0 - x_i)\Big), \quad i=1,\ldots,N,
\end{align}
while the leader dynamics are given by the equation
\begin{align} \label{leader dymamic}
 \dot{x}_0=A(\rho(t))x_0 +  B_2\sum\limits_{k\colon \bar d_k=1}\varphi_{0k} (t,
 x_k-x_0).
\end{align}
Unlike agents $i$, $i=1,\ldots
  N$, the leader is not controlled, i.e., $u_0\equiv 0$. On the contrary, all other agents will be
  controlled to track node 0.

\begin{remark}
In this paper, no \emph{a priori} stability assumptions are made about
the matrices $A(\rho)$. This is in contrast to, for example, the
synchronization problem considered in \cite{Scardovia2009}, where the
dynamics of the leader were required to be stable or marginally stable.
In the sequel, however,
certain conditions, in an LMI form, will be imposed on the
coefficients of the leader and followers dynamics. While we formally make
no special provisions regarding these coefficients, except for
the feasibility of these LMI conditions, for such LMIs to be feasible
one can reasonably expect the overall system dynamics to have
certain collective stabilizability and detectability properties; see
Remark~\ref{feasibility-remark} below.
\end{remark}

Define tracking
performance associated with the control input $u(\cdot)=[u_1(\cdot)'~\ldots~u_N(\cdot)']'$ as
\begin{align} \label{cost function}
 \mathcal {J}(u) &=\sum\limits_{i=1}^{N} \int_{0}^{\infty}
 \left((x_0-x_i)'Q (x_0-x_i) + u_i'R u_i\right) dt,
\end{align}
where $Q=Q'>0$ and $R=R'>0$ are given weighting matrices.
In this paper we are concerned with the following problem.
\begin{problem}
\label{prob 1}
For each follower $i$, find gain schedule functions $K_i\colon \Gamma\to
\Re^{p\times n}$ so that the control protocols of the form
\begin{equation} \label{controller}
 u_i=-K_i(\rho(t))\left\{\sum\limits_{j\in S^c_i}(x_j-x_i) + g_i(x_0-x_i)\right\},
\end{equation}
ensure a bounded worst-case tracking performance,
\begin{equation} \label{cost function 1}
\sup \limits_{\Xi}\mathcal {J}(u) < \mathrm{const},
\end{equation}
where $\mathrm{const}$ denotes a constant which can depend on $x_0(0)$ and
$x_i(0)$.
\end{problem}

\section{The Main Results}
\label{main}

In this section, we first revisit the leader
follower tracking control problem for multi-agent systems with fixed parameters
considered in \cite{[Cheng2013b]} to obtain a sufficient condition for the
design of a control protocol for a more general class of
systems involving coupling between the leader and the agents. This
sufficient condition will then be applied to develop an
interpolation technique to obtain a continuous gain scheduling
tracking control protocol for the system (\ref{all agents dymamic}). The interpolation technique is the main result of this paper.

\subsection{Leader follower control for fixed parameter systems}

We now revisit the results of \cite{[Cheng2013b]}. This revision is
  prompted by a more general structure of the system (\ref{all agents
    dymamic}) which allows for physical connections between the leader and
  the followers.  As a result, the form of the control protocol is somewhat
  different here in that the resulting controller gains depend on
  $i$.

Consider a fixed-parameter version of the system (\ref{all agents dymamic})
described by the equation
\begin{align} \label{all agents dymamic.fixed}
\dot{x}_i=&A(\rho)x_i+\beta\xi_i(t, x_i) +B_1u_i+ B_2\sum\limits_{j\in N^\varphi_i}
\varphi_{ij} (t, x_j-x_i), \quad i=0,\ldots, N,
\end{align}
where $\rho\in \Gamma $ is fixed, and $\beta$ is a positive
constant. Compared to (\ref{all
  agents dymamic}), the system (\ref{all agents dymamic.fixed}) includes an
additional uncertainty element $\xi_i(t,x_i)$, which satisfies the
following constraint
\begin{equation}\label{present.constraint}
\|\xi_0(t,x)-\xi_i(t,y)\|^2\le \|x-y\|^2, \quad \forall x,y\in \Re^n.
\end{equation}
In the sequel, we will show that when the difference $A(\rho(t))-A(\rho)$
is sufficiently small, the system  (\ref{all
  agents dymamic}) can be represented
as the system (\ref{all agents dymamic.fixed})  subject to
(\ref{present.constraint}).

We now derive a distributed protocol of the form
(\ref{controller}) under which the fixed-parameter uncertain system
(\ref{all agents dymamic.fixed}) satisfies the performance requirement
(\ref{cost function 1}).

Define the leader tracking error vectors as $e_i=x_0-x_i$, $i=1, \ldots, N$. Dynamics of the variable $e_i$ satisfy the equation
 \begin{align} \label{fixed simu error dymamic}
 \dot{e}_i=&A(\rho)e_i-B_1u_i + \beta \tilde \xi_i(t) - B_2
 \sum\limits_{k\colon \bar d_k=1} \varphi_{0k}(t,e_k)\nonumber \\
 &- B_2 \sum\limits_{j\in S^\varphi_i} \big(\varphi_{ij}(t,e_i) - \varphi_{ij}(t,e_j)\big)- B_2 d_i \varphi_{i0}(t,e_i),
  \end{align}
 where $\tilde\xi_i(t)\triangleq \xi_0(t,x_0(t))-\xi_i(t,x_i(t))$. It
 follows from
 (\ref{present.constraint}) that $\|\tilde \xi_i\|^2\le \|e_i\|^2$ for all
 $t\ge 0$.

For node $i$ of the subgraph $\mathcal {G}^\varphi_0$, introduce matrices $\hat
C_i=[C_{ij_1}' \ldots
C_{ij_{\kappa_i}}']'$, $\bar C_i=[C_{r_1i}' \ldots
C_{r_{\chi_i}i}']'$, where $j_1,\ldots, j_{\kappa_i}$ are the elements of the
neighborhood  set $S^\varphi_i$, and $r_1,\ldots,r_{\chi_i}$ are the nodes with the
property $(i, r_\iota)\in\mathcal{E}^\varphi_0$; $\kappa_i$ and $\chi_i$ are,
respectively, the in-degree and the out-degree of node $i$ in the graph
$\mathcal {G}^\varphi_0$. Also, let $\bar Q=(\sigma^2/\hat \lambda) Q $, where
$\sigma$, $\hat\lambda$ are the constants defined
in~(\ref{sigma.lambdahat}).

In order to formulate the extension of Theorem~1 in \cite{[Cheng2013b]},
with each node $i$, $i=1,\dots, N$, we associate a collection of positive
constants $\nu_{ij}$, $\mu_{ij}$, $j\in S^\varphi_i$, $\pi_{i}$, $\nu_{i0}$ (only
for those nodes $i$ for which $d_i=1$), and  $\mu_{0i}$ (only
for those nodes $i$ for which $\bar d_i=1$). Also, let $Y=Y'>0$ be an $n\times n$ matrix. Using these constants and the matrix, for each node $i$
introduce a matrix $\Pi_i$ defined depending on $d_i$, $\bar
d_i$ as follows.

\emph{Case 1. $d_i\neq 0$ and $\bar d_i\neq 0$. } For each such node $i$, define the
matrix $\Pi_i$ as
\begin{eqnarray}
\scriptsize
\Pi_i=\left[\begin{array}{ccccc|cc}
Z_i     & Y \bar Q ^{1/2}  & Y\hat C_i'&  Y\bar C_i' &  Y  &  Y C_{i0}' &  Y C_{0i}' \\
 \bar Q^{1/2}Y & -I & 0 & 0 & 0 & 0 & 0 \\
\hat C_iY & 0 & -\Phi_i & 0 & 0 & 0 & 0  \\
\bar C_i Y & 0 & 0  &  -\Omega_i& 0 & 0 & 0 \\
 Y  & 0 & 0  &  0& -\Lambda_i & 0 & 0\\[2pt] \hline
 C_{i0} Y & 0 & 0  &  0 & 0 & -\frac{1}{\nu_{i0}}I& 0  \\
C_{0i} Y & 0 & 0  &  0& 0 & 0 & -\frac{1}{N \mu_{0i}}I
\end{array}\right],
\label{Pi.d_i=1}
\end{eqnarray}
where
\begin{align}
\Phi_i=&\mathrm{diag}[\frac{1}{\nu_{ij}}I, j\in S^\varphi_i],\quad \Omega_i=\mathrm{diag}[\frac{1}{\mu_{ji}}I, j\colon i\in S^\varphi_j],\quad \Lambda_i=\frac{1}{\pi_{i}}I,  \nonumber \\
Z_i=&A(\rho)Y + Y A(\rho)' - B_1 R^{-1} B_1'+ \frac{1}{\pi_{i}}\beta^2I \nonumber \\
 &+ \big(\sum \limits_{j\in S^\varphi_i}(\frac{1}{\nu_{ij}}+ \frac{1}{\mu_{ij}}) + \frac{1}{\nu_{i0}}+ \sum \limits_{k\colon \bar d_k=1} \frac{1}{\mu_{0k}}\big) B_2B_2'. \quad
\label{Z_i.d_i=1}
\end{align}

\emph{Case 2. $d_i= 0$ and $\bar d_i\neq 0$. } For each such node $i$, the constant
$\nu_{i0}$ is not defined. Accordingly, we define the
matrix $\Pi_i$ by removing the second last column and row from the matrix
in (\ref{Pi.d_i=1}):
\begin{eqnarray}
\Pi_i=\left[\begin{array}{ccccc|c}
Z_i     & Y \bar Q ^{1/2}  & Y\hat C_i'&  Y\bar C_i' &  Y &  Y C_{0i}' \\
 \bar Q^{1/2}Y & -I & 0 & 0 & 0 & 0\\
\hat C_iY & 0 & \Phi_i & 0 & 0 & 0 \\
\bar C_i Y & 0 & 0  &  \Omega_i & 0& 0  \\
 Y  & 0 & 0  &  0& \Lambda_i & 0\\
 \hline
 C_{0i} Y & 0 & 0  &  0& 0  & -\frac{1}{N \mu_{0i}}I
\end{array}\right],
\label{Pi.d_i=0}
\end{eqnarray}
where the matrix $Z_i$ is modified to be
\begin{eqnarray}
Z_i=A(\rho)Y + Y A(\rho)' - B_1 R^{-1} B_1'+ \frac{1}{\pi_{i}}\beta^2I + \big(\sum \limits_{j\in S^\varphi_i}(\frac{1}{\nu_{ij}}+ \frac{1}{\mu_{ij}}) +
 \sum \limits_{k\colon \bar d_k=1} \frac{1}{\mu_{0k}} \big) B_2B_2'.
\label{Z_i.d_i=0}
\end{eqnarray}

\emph{Case 3. $d_i\neq 0$ and $\bar d_i= 0$. } For each such node the constant
$\mu_{0i}$ is not defined, hence the corresponding matrix $\Pi_i$ will be
defined by removing the last column and row from the matrix in
(\ref{Pi.d_i=1}):
\begin{eqnarray}
\Pi_i=\left[\begin{array}{ccccc|c}
Z_i     & Y \bar Q ^{1/2}  & Y\hat C_i'&  Y\bar C_i' &  Y  &  Y C_{i0}' \\
 \bar Q^{1/2}Y & -I & 0 & 0 & 0 & 0 \\
\hat C_iY & 0 & -\Phi_i & 0 & 0 & 0 \\
\bar C_i Y & 0 & 0  &  -\Omega_i& 0 & 0 \\
 Y  & 0 & 0  &  0& -\Lambda_i & 0 \\[2pt] \hline
 C_{i0} Y & 0 & 0  &  0 & 0 & -\frac{1}{\nu_{i0}}I
\end{array}\right],
\label{Pi.d_i=1.bar d_i=0}
\end{eqnarray}
The matrix $Z_i$ is the same as in (\ref{Z_i.d_i=1}).

\emph{Case 4. $d_i= 0$ and $\bar d_i= 0$. } In this case, both $\nu_{i0}$ and
$\mu_{0i}$ are not defined, and the corresponding matrix $\Pi_i$ is
defined by removing two last columns and rows from the matrix in (\ref{Pi.d_i=1}):
\begin{eqnarray}
\Pi_i=\left[\begin{array}{ccccc}
Z_i     & Y \bar Q ^{1/2}  & Y\hat C_i'&  Y\bar C_i' &  Y   \\
 \bar Q^{1/2}Y & -I & 0 & 0 & 0 \\
\hat C_iY & 0 & -\Phi_i & 0 & 0  \\
\bar C_i Y & 0 & 0  &  -\Omega_i& 0 \\
 Y  & 0 & 0  &  0& -\Lambda_i
\end{array}\right],
\label{Pi.d_i=0.bar d_i=0}
\end{eqnarray}
The matrix $Z_i$ is the same as in (\ref{Z_i.d_i=0}).

The following theorem is an extension of Theorem 1 in \cite{[Cheng2013b]}.

\begin{theorem}
\label{Theorem 1}
Under Assumption 1, let a matrix $ Y= Y'> 0$, $Y \in \Re ^{n\times n}$,
constants $\nu_{ij}>0$, $\mu_{ij}>0$, $\pi_{i} > 0$, $j\in S^\varphi_i$,
$i=1,\ldots, N$, and constants $\nu_{i0}>0$ (for those $i$
with $d_i\neq 0$), $\mu_{0i}>0$ (for those $i$ with $\bar d_i\neq 0$) exist
such that the following LMIs are satisfied simultaneously
\begin{eqnarray}
\Pi_i<0, \quad i=1,\ldots, N.
\label{LMI Dircted}
\end{eqnarray}
Then the control protocol (\ref{controller}) with
\begin{align}
\label{fixed control gain}
K_i(\rho)=- (\vartheta_i\sigma)^{-1}R^{-1}B_1'Y^{-1}
\end{align}
 solves the leader follower tracking
  control problem for the fixed parameter system (\ref{all agents
    dymamic.fixed}).
Furthermore, this
protocol guarantees the following performance bound
 \begin{equation} \label{cost function TH1}
 \sup_{\Xi}\mathcal {J}(u) \le \frac{\hat \lambda}{\sigma^2}\sum_{i=1}^{N} (x_0(0)-x_i(0))'Y^{-1}(x_0(0)-x_i(0))
\end{equation}
for all uncertainties $\xi_i$ for which (\ref{present.constraint}) holds.
\end{theorem}

\begin{remark}\label{feasibility-remark}
  The feasibility of the LMIs (\ref{LMI Dircted}) in Theorem~\ref{Theorem 1}
  is a sufficient condition for the existence of a control scheme
  to guarantee the performance bound (\ref{cost function TH1}). The LMIs are
  numerically tractable and can be solved
  using the existing software. Additionally, using the standard tools of
  the $H_\infty$ control theory, such as the Strict Bounded Real Lemma and
  the associated Riccati equation, the feasibility of the LMIs (\ref{LMI
    Dircted}) can be related to collective stabilizability and/or detectability
  properties of the coefficients of the interconnected system (\ref{all
    agents dymamic.fixed}). Since this system consists of identical agents,
  one can expect that each agent must necessarily have corresponding
  stabilizability and detectability properties for these collective
  properties to hold; cf.~\cite{U7b-journal}.
\end{remark}

\emph{Proof: }
Using the Schur complement, each LMI (\ref{LMI Dircted}) can be transformed
into the following Riccati inequality:
\begin{align}
\label{ARI}
&A(\rho)Y +Y  A(\rho)' - B_1 R^{-1} B_1' + \big(\sum \limits_{j\in
    S^\varphi_i}(\frac{1}{\nu_{ij }}+ \frac{1}{\mu_{ij }}) + \sum \limits_{k\colon \bar d_k=1} \frac{1}{\mu_{0k}} \big) B_2B_2' + \frac{1}{\pi_{i}}\beta^2I  \nonumber \\
& +Y \big( \bar Q + \sum \limits_{j\in S^\varphi_i} \nu_{ij }C'_{ij}C_{ij} + \sum\limits_{j\colon i \in S^\varphi_j}\mu_{ji }C'_{ji}C_{ji} +\pi_{i} I \big)Y \nonumber \\
& + \left(\frac{1}{\nu_{i0 }} B_2B_2' + \nu_{i0}YC'_{i0}C_{i0}Y\right)
\nonumber \\
& \qquad\qquad\qquad \mbox{(this term is present only if $d_i=1$)} \nonumber \\
& + N \mu_{0i }YC'_{0i}C_{0i}Y  < 0 \\
& \qquad\qquad\qquad \mbox{(this term is present only if $\bar d_i=1$)}. \nonumber
\end{align}
Note that the last and the second last lines in the Riccati inequality
(\ref{ARI}) are present only for
those nodes $i$ for which $d_i\neq 0$ and/or $\bar d_i\neq 0$, respectively.

After pre- and post-multiplying (\ref{ARI}) by $Y^{-1}$, and then using
the expression (\ref{fixed control gain}) for $K_i(\rho)$ in the resulting
inequality, we obtain
\begin{align}
\label{ARI in2}
& Y^{-1} (A(\rho)+ \sigma \vartheta_i B_1K_i(\rho)) + (A(\rho)+ \sigma \vartheta_i B_1K_i(\rho))'Y^{-1} + \sum \limits_{j\in S^\varphi_i} \nu_{ij }C'_{ij}C_{ij} + \sum\limits_{j\colon i \in S^\varphi_j}\mu_{ji }C'_{ji}C_{ji}  \nonumber\\
&  + Y^{-1}\Big( \big( \sum \limits_{j\in S^\varphi_i}(\frac{1}{\nu_{ij }}+ \frac{1}{\mu_{ij }}) + \sum \limits_{k\colon \bar d_k=1} \frac{1}{\mu_{0k}}\big)B_2B_2' + \frac{1}{\pi_{i }}\beta^2I \Big)Y^{-1} +  Y^{-1}B_1 R^{-1} B_1'Y^{-1} + \bar Q +\pi_{i } I \nonumber \\
& +  \left(\frac{1}{\nu_{i0 }} Y^{-1} B_2B_2'Y^{-1} + \nu_{i0}C'_{i0}C_{i0}\right)
\nonumber \\
& \qquad\qquad\qquad \mbox{(this term is present only if $d_i=1$)} \nonumber \\
& + N \mu_{0i}C'_{0i}C_{0i}  < 0\\
& \qquad\qquad\qquad \mbox{(this term is present only if $\bar d_i=1$)}. \nonumber
\end{align}

Define $e=[e'_1, \ldots, e'_N]'$ and consider the following Lyapunov
function candidate for the interconnected system consisting of the
subsystems (\ref{fixed simu error dymamic}):
\begin{align}
V(e)= \sum \limits_{i=1}^N e_i' Y^{-1}e_i.
\end{align}
Then
\begin{align}
\label{lyapunov equation1.1}
 \frac{d V(e)}{dt}=& \sum \limits_{i=1}^N 2 e_i' Y^{-1}\Big(A(\rho)e_i +  B_1K_i (\rho)(\sum \limits_{j\in S^c_i}(e_i-e_j)+g_ie_i) \Big) \nonumber\\
&  - 2\sum\limits_{i=1}^N  \sum\limits_{k\colon \bar d_k=1} e_i' Y^{-1}B_2\varphi_{0k}(t,e_k) + 2\sum \limits_{i=1}^N e_i' Y^{-1} \beta \tilde\xi_i(t, e_i) -  2\sum\limits_{i=1}^Nd_i e_i' Y^{-1}B_2 \varphi_{i0}(t,e_i) \nonumber\\
&- 2\sum \limits_{i=1}^N \sum\limits_{j\in S^\varphi_i} e_i' Y^{-1}B_2  \varphi_{ij} (t, e_i)+ 2\sum \limits_{i=1}^N \sum\limits_{j\in S^\varphi_i}e_i' Y^{-1}B_2  \varphi_{ij} (t, e_j).
\end{align}
Note the following inequality:
\begin{align}
\label{one trick1}
&\sum \limits_{i=1}^N 2  e_i'Y^{-1} B_1K_i (\rho)(\sum \limits_{j\in S^c_i}(e_i-e_j)+g_ie_i) \nonumber\\
 &=-2e'(\Theta(\mathcal {L}^c_0+G)\otimes (Y^{-1}B_1(\sigma R)^{-1}B_1'Y^{-1}))e \nonumber\\
&= -y'\big((\Theta(\mathcal {L}^c_0+G)+ (\mathcal {L}^c_0+G)'\Theta)\otimes I_p\big)y \nonumber\\
&\leq -2 \sigma y'\big(I_N\otimes I_p\big)y \nonumber\\
&= -2\sigma e'\big(I_N \otimes Y^{-1}B_1 (\sigma R)^{-1}B_1'Y^{-1}\big)e \nonumber\\
&=-2\sum \limits_{i=1}^N e_i' Y^{-1}B_1 R ^{-1}B_1'Y^{-1} e_i,
\end{align}
where $y=(I_N \otimes (\sigma R)^{-1/2}B_1'Y^{-1}) e$.

From (\ref{lyapunov equation1.1}) and (\ref{one trick1}), one has
\begin{align}
\label{lyapunov equation1112}
\frac{d V(e)}{dt}\leq& \sum \limits_{i=1}^N 2 e_i' Y^{-1}\Big( A(\rho)+ \sigma \vartheta_i B_1K_i(\rho) \Big)e_i  + 2\sum \limits_{i=1}^N e_i' Y^{-1} \beta \tilde \xi_i(t, e_i)\nonumber\\
& -2\sum \limits_{i=1}^N \sum\limits_{j\in S^\varphi_i} e_i' Y^{-1}B_2  \varphi_{ij} (t, e_i)- 2\sum\limits_{i=1}^N\sum\limits_{k\colon \bar d_k=1} e_i' Y^{-1}B_2\varphi_{0k}(t,e_k) \nonumber\\
&-  2\sum\limits_{i=1}^N d_i e_i' Y^{-1}B_2 \varphi_{i0}(t,e_i) +2\sum \limits_{i=1}^N \sum\limits_{j\in S^\varphi_i}e_i' Y^{-1}B_2  \varphi_{ij} (t, e_j).
\end{align}

Using the Riccati inequality (\ref{ARI in2}), it follows from (\ref{lyapunov equation1112}) that
\begin{align}
\label{lyapunov equation33}
\frac{d V(e)}{dt} \leq & - \sum \limits_{i=1}^N e_i' \Big(Y^{-1}B_1 R^{-1} B_1'Y^{-1} + \bar Q  + \sum \limits_{j\in S^\varphi_i} \nu_{ij }C'_{ij}C_{ij} + \sum\limits_{j\colon i \in S^\varphi_j}\mu_{ji }C'_{ji}C_{ji} \nonumber\\
&  +\pi_{i } I + Y^{-1}\big((\sum \limits_{j\in S^\varphi_i}(\frac{1}{\nu_{ij }}+ \frac{1}{\mu_{ij }}) +
\sum \limits_{k\colon \bar d_k=1} \frac{1}{\mu_{0k}})B_2B_2' + \frac{1}{\pi_{i }}\beta^2I \big)Y^{-1} \Big)e_i  \nonumber\\
& + \sum \limits_{i\colon d_i=1}^N e_i' \Big(\frac{1}{\nu_{i0}} Y^{-1}B_2B_2'Y^{-1} + \nu_{i0}C'_{i0}C_{i0}\Big)e_i + N \sum \limits_{i\colon \bar d_i=1}^N e_i' \mu_{0i}C'_{0i}C_{0i}e_i \nonumber\\
& + 2\sum \limits_{i=1}^N e_i' Y^{-1} \beta \tilde \xi_i(t,
e_i) -2\sum \limits_{i=1}^N \sum\limits_{j\in S^\varphi_i} e_i' Y^{-1}B_2  \varphi_{ij}
(t, e_i) - 2\sum\limits_{i=1}^N d_i e_i' Y^{-1}B_2 \varphi_{i0}(t,e_i) \nonumber\\
& -2\sum\limits_{i=1}^N \sum\limits_{k\colon \bar d_k=1} e_i' Y^{-1}B_2\varphi_{0k}(t,e_k) + 2\sum \limits_{i=1}^N \sum\limits_{j\in S^\varphi_i}e_i' Y^{-1}B_2  \varphi_{ij} (t, e_j).
\end{align}

Using the following identities,
\begin{align*}
\sum\limits_{i=1}^{N} \sum\limits_{j\in S^\varphi_i} \mu_{ij }e'_j C'_{ij}C_{ij}e_j =& \sum\limits_{i=1}^{N} \sum\limits_{j\colon i \in S^\varphi_j} \mu_{ji }e'_i C'_{ji}C_{ji}e_i,  \\
N \sum \limits_{i\colon \bar d_i=1}^N e_i' \mu_{0i}C'_{0i}C_{0i}e_i '=&N \sum \limits_{k\colon \bar d_k=1}^N e_k' \mu_{0k}C'_{0k}C_{0k}e_k \\
\sum\limits_{i=1}^N d_i e_i' Y^{-1}B_2 \varphi_{i0}(t,e_i)=& \sum\limits_{i \colon d_i=1} e_i' Y^{-1}B_2 \varphi_{i0}(t,e_i),
\end{align*}
and completing the squares, one has
\begin{align}
\label{lyapunov equation44}
&\frac{d V(e)}{dt} \leq  - \sum \limits_{i=1}^{N} e_i' \Big(Y^{-1}B_1 R^{-1} B_1'Y^{-1} + \bar Q \Big)e_i\nonumber \\
&-\sum \limits_{i=1}^{N} \|\frac{\beta}{\sqrt {\pi_{i }}}Y^{-1}e_i - \sqrt {\pi_{i }} \tilde \xi_i(t, e_i)\|^2 + \sum \limits_{i=1}^{N}\pi_{i } (\|\tilde \xi_i(t, e_i)\|^2-\|e_i\|^2 )\nonumber \\
&-\sum \limits_{i=1}^{N} \sum\limits_{j\in S^\varphi_i} \|\frac{1}{\sqrt {\nu_{ij }}}B_2'Y^{-1}e_i + {\sqrt {\nu_{ij }}} \varphi_{ij} (t, e_i)\|^2 + \sum \limits_{i=1}^{N}\sum\limits_{j\in S^\varphi_i} \nu_{ij } (\|\varphi_{ij}
(t, e_i)\|^2-\|C_{ij}e_i\|^2 )\nonumber\\
&- \sum \limits_{i=1}^{N}\sum\limits_{j\in S^\varphi_i}\| \frac{1}{\sqrt {\mu_{ij }}}B_2'Y^{-1}e_i - {\sqrt {\mu_{ij }}}\varphi_{ij} (t, e_j)\|^2 + \sum \limits_{i=1}^{N}\sum\limits_{j\in S^\varphi_i} \mu_{ij }(\| \varphi_{ij} (t, e_j)\|^2 - \|C_{ij}e_j\|^2) \nonumber \\
&-\sum \limits_{i\colon d_i=1} \|\frac{1}{\sqrt {\nu_{i0}}}B_2'Y^{-1}e_i + {\sqrt {\nu_{i0}}} \varphi_{i0} (t, e_i)\|^2 + \sum \limits_{i\colon d_i=1}\nu_{i0} (\|\varphi_{i0} (t, e_i)\|^2-\|C_{i0}e_i\|^2 ) \nonumber\\
&- \sum \limits_{i=1}^{N} \sum\limits_{k\colon \bar d_k=1} \|\frac{1}{\sqrt {\mu_{0k}}}B_2'Y^{-1}e_i + {\sqrt {\mu_{0k}}} \varphi_{0k} (t, e_k)\|^2 + N \sum\limits_{k\colon \bar d_k=1} \mu_{0k} (\|\varphi_{0k} (t, e_k)\|^2-\|C_{0k}e_k\|^2 ).
\end{align}

According to the norm-bounded condition (\ref{nb}),  from
(\ref{lyapunov equation44}) we have
\begin{align}
\label{lyapunov equation441}
\int_{0}^{t}\frac{d V(e)}{dt}dt  \leq -  \sum \limits_{i=1}^{N}  \int_{0}^{t}e_i' \Big( Y^{-1}B_1 R^{-1} B_1'Y^{-1} + \bar Q \Big)e_i  dt.
\end{align}

Since $V (e(t)) \ge 0$, then (\ref{lyapunov equation441}) implies
\begin{align}
\label{lyapunov equation44-1}
 \sum \limits_{i=1}^{N}  \int_{0}^{t}e_i' \Big( Y^{-1}B_1 R^{-1} B_1'Y^{-1} + \bar Q \Big)e_i  dt  \leq V(e(0)).
\end{align}

The expression on the right hand side of the above inequality is independent of $t$. Letting $t\rightarrow \infty$  leads to
\begin{align}
 \sum \limits_{i=1}^{N}  \int_{0}^\infty e_i' \Big( Y^{-1}B_1 R^{-1} B_1'Y^{-1} + \bar Q \Big)e_i  dt \leq V(e(0)).
\end{align}
Using (\ref{cost function}) and (\ref{controller}), we have
\begin{align}
 \mathcal {J}(u)=&\sum\limits_{i=1}^{N} \int_{0}^{\infty} \Big(e_i'Q  e_i + u_i'R u_i\Big) dt  \nonumber \\
 =&\int_{0}^{\infty} \Big(e'(I_N \otimes Q ) e + e'\big ((\mathcal{L}^c_0+G)'\Theta^2(\mathcal{L}^c_0+G)\otimes \frac{1}{\sigma^2} Y^{-1}B_1 R^{-1} B_1'Y^{-1}\big )e\Big) dt \nonumber\\
 \leq& \int_{0}^{\infty} \Big(e'(I_N \otimes Q ) e  + e' \big ( I_N\otimes \frac{\hat \lambda}{\sigma^2} Y^{-1}B_1 R^{-1} B_1'Y^{-1}\big ) e\Big) dt  \nonumber\\
 =&\sum \limits_{i=1}^{N}  \int_{0}^\infty e_i' \Big( \frac{\hat \lambda}{\sigma^2}Y^{-1}B_1 R^{-1} B_1'Y^{-1} + Q \Big)e_i  dt.
\end{align}
Since $Q=\frac{\hat \lambda}{\sigma^2} \bar Q$, it follows from (\ref{lyapunov equation44-1}) that
\begin{align}
 \mathcal {J}(u)&\leq \frac{\hat \lambda}{\sigma^2} \sum \limits_{i=1}^{N}  \int_{0}^\infty e_i' \Big( Y^{-1}B_1 R^{-1} B_1'Y^{-1} + \bar Q \Big)e_i \leq \frac{\hat \lambda}{\sigma^2}  \sum_{i=1}^{N} e_i'(0)Y^{-1}e_i(0).
 \end{align}
It implies that the control protocol (\ref{controller}) with $K_i(\rho)$ defined in (\ref{fixed control gain}) solves leader following tracking control problem, and also guarantees
  the performance bound (\ref{cost function TH1}).
\hfill$\Box$

Theorem~\ref{Theorem 1} can be applied to obtain a leader-follower tracking
protocol (\ref{controller}) for the system (\ref{all agents dymamic}) if
parameter variations of systems are sufficiently small.

Suppose there exists $\rho_0\in \Gamma$ and $\beta>0$ such that
\begin{align}
\label{A,B}
(A(\rho(t))-A({\rho_0}))'(A(\rho(t))-A({\rho_0}))\leq \beta^2 I, \quad \forall ~t \in [0,\infty).
\end{align}
and define $\xi_i(t, x_i):=\frac{1}{\beta}[A(\rho(t))-A({\rho_0})]x_i$.
This allows us
to regard small variations of the matrix $A(\cdot)$ as perturbations. The
fixed-parameter system (\ref{all agents dymamic.fixed}) with
$\rho\equiv\rho_0$ captures this type of perturbations.
Then the following result follows from Theorem~\ref{Theorem 1}.

\begin{corollary}
\label{small variations}
Under Assumption \ref{graph assumption}, if the LMIs (\ref{LMI Dircted})
with $\rho=\rho_0$ and $Y=Y_{\rho_0}$ are satisfied simultaneously,
then the control protocol (\ref{controller}) with
\begin{align}
K_i(\rho_0)=- (\vartheta_i\sigma)^{-1} R^{-1}B_1'Y_{\rho_0}^{-1}
\end{align}
 solves the leader following tracking
  control problem for the parameter varying system (\ref{all agents
    dymamic}) under small variations of $\rho(t)$ for which condition (\ref{A,B}) holds
  for all $t>0$. Furthermore, this
protocol guarantees the following performance bound
 \begin{equation} \label{cost function TH1 small}
 \sup_{\Xi}\mathcal {J}(u) \le \frac{\hat \lambda}{\sigma^2} \sum_{i=1}^{N} (x_0(0)-x_i(0))'Y_{\rho_0}^{-1}(x_0(0)-x_i(0)).
 \end{equation}
\end{corollary}

\subsection{Design of a continuous protocol schedule}

The result of Corollary~\ref{small variations} only holds under assumption
that variations of the matrix $A(\rho(\cdot))$ are sufficiently small to
satisfy (\ref{A,B}). In general, it may be difficult to satisfy (\ref{A,B}) using a single $\rho_0$, or the
corresponding LMIs of Corollary~\ref{small variations} may not be
feasible. In order to address this situation, we propose a gain scheduling
approach.

Consider a set of design points
$\Gamma_\ell:=\{\rho_s, s=1, \ldots, M\} \subset \Gamma $ and a collection of positive constants $\beta_s$ chosen so
that for any $\rho \in \Gamma$ there exists
at least one point $\rho_s$ with the property
\begin{align}
\label{designpoints}
(A(\rho)-A({\rho_s}))'(A(\rho)-A({\rho_s}))\leq \beta_s^2 I
\end{align}
and that the LMIs (\ref{LMI Dircted}) with $\rho=\rho_s$ are feasible.

Let $U_s$ be the largest connected neighborhood of the design
point $\rho_s \in \Gamma_\ell$ such that (\ref{designpoints}) holds if
$\rho=\rho(t)\in U_s$. This allows a protocol (\ref{controller}) to be
scheduled for each value $\rho$ on the trajectory $\rho(t)$,
by associating with every $\rho(t)$ the protocol computed using
Theorem~\ref{Theorem 1} for one of the
indexes $s \in \{s\colon \rho(t) \in U_s \}$.
However, when applied to the parameter-varying
system (\ref{all agents dymamic}), the gains of such a protocol may become
discontinuous at the time instant when $\rho(t)$ is switching between
different sets $U_s$. To overcome this issue, the continuous interpolation
technique proposed in \cite{Stilwell1999,[Yoon2007]} is used in this paper to
obtain a continuous consensus control protocol; also
see~\cite{U8}.

Consider an arbitrary fixed $\rho \in \Gamma$, and the collection of
constants $\beta_s$ and grid points $\Gamma_\ell$. Select $s$ such that
$\rho \in U_s$, and let $({\pi}_{i,{\rho_s}}, {\nu}_{ij,{\rho_s}},
{\mu}_{ij,{\rho_s}}, Y_{\rho_s},| {\nu}_{i0,{\rho_s}},
{\mu}_{0i,{\rho_s}})$, $j\in S^\varphi_i$, $i= 1, \ldots, N$, be
a feasible solution to the LMIs (\ref{LMI Dircted}). Recall that
${\nu}_{i0,{\rho_s}}$ and ${\mu}_{0i,{\rho_s}}$ are only defined for those
nodes $i$ with $d_i \neq 0$ and $\bar d_i \neq 0$, respectively. It is
straightforward to show that this collection of positive constants and the
matrix is also a feasible solution to the following reduced coupled LMIs
\begin{align}
\label{LMI Dircted uncertain}
\Upsilon_i<0, \quad i=1, \ldots, N,
\end{align}
where the matrix $\Upsilon_i$ is defined as follows.
For those nodes $i$, for which $d_i\neq 0$ and $\bar d_i\neq 0$,
$\Upsilon_i$ is defined by removing the fifth column and row from the
matrix $\Pi_i$ defined in (\ref{Pi.d_i=1})  and replacing the matrix $Z_i$ in
(\ref{Pi.d_i=1}) with the following matrix:
\begin{align*}
Z_i=A(\rho)Y+Y A(\rho)'  - B_1 R^{-1} B_1' + \Big(\sum \limits_{j\in S^\varphi_i}(\frac{1}{\nu_{ij}}+ \frac{1}{\mu_{ij}}) +
\frac{1}{\nu_{i0}} + \sum \limits_{k\colon \bar d_k=1} \frac{1}{\mu_{0k}}
\Big)B_2B_2'.
\end{align*}
where the matrices $\Phi_i$ and $\Omega_i$ are as defined previously. That
is, the new matrix $Z_i$ is obtained from the matrix in (\ref{Z_i.d_i=1})
by subtracting the term $\frac{1}{\pi_i}\beta^2I$. This results in the
matrix $\Upsilon_i$ defined as
\begin{align}
\label{LMI coupled with leader}
\Upsilon_i=\left[
\begin{array}{cccc|cc}
Z_i     & Y \bar Q ^{1/2}  & Y\hat C_i'&  Y \bar C_i'& Y C_{i0}'&  Y C_{0i}'\\
 \bar Q^{1/2}Y  & -I & 0 & 0 & 0 & 0 \\
\hat C_iY  & 0 & -\Phi_i & 0 & 0 & 0 \\
\bar C_i Y  & 0 & 0  &  -\Omega_i& 0 & 0\\
\hline
 C_{i0} Y  & 0 & 0  &   0& -\frac{1}{\nu_{i0}}I & 0\\
 C_{0i} Y  & 0 & 0  &  0& 0 & -\frac{1}{N\mu_{0i}}I
\end{array}\right],
\end{align}
For three other cases, the matrix $\Upsilon_i$ is defined in the same
fashion. First, the fifth column and row are removed from the
matrix $\Pi_i$ defined in (\ref{Pi.d_i=0}), (\ref{Pi.d_i=1.bar d_i=0}), or
(\ref{Pi.d_i=0.bar d_i=0}), respectively. Next, the matrix $Z_i$ is
redefined by subtracting
$\frac{1}{\pi_i}\beta^2I$ from the corresponding matrix in
(\ref{Z_i.d_i=1}) or (\ref{Z_i.d_i=0}), as appropriate.

Now consider the uncertain fixed parameter system (\ref{fixed simu error
  dymamic}), and assume $\rho \in U_s \cap U_{s+1}$. Then we conclude that
both collections $({\nu}_{ij,{\rho_s}}, {\mu}_{ij,{\rho_s}}, Y_{\rho_s},|
{\nu}_{i0,{\rho_s}}, {\mu}_{0i,{\rho_s}})$ and $({\nu}_{ij,{\rho_{s+1}}},
{\mu}_{ij,{\rho_{s+1}}}, Y_{\rho_{s+1}},| {\nu}_{i0,{\rho_{s+1}}},
{\mu}_{0i,{\rho_{s+1}}})$, $j\in S^\varphi_i$, $i= 1, \ldots, N$,
are feasible solutions to the coupled LMIs (\ref{LMI Dircted
  uncertain}).

This allows us to construct interpolated
feasible solutions to (\ref{LMI Dircted uncertain}) as follows.

For a $\gamma \in [0, 1]$, define
\begin{align}
Y_{\gamma} &= \gamma Y_{\rho_s} + (1-\gamma)Y_{\rho_{s+1}},  \\
 {\nu}_{ij,\gamma} &= [\gamma {\nu}_{ij,{\rho_s}}^{-1} +(1-\gamma){\nu}^{-1}_{ij,{\rho_{s+1}}}]^{-1},\\
 {\mu}_{ij,\gamma} &= [\gamma{\mu}_{ij,{\rho_s}}^{-1}
 +(1-\gamma){\mu}^{-1}_{ij,{\rho_{s+1}}}]^{-1}.
\end{align}
Also, for nodes $i$ with $d_i=1$, define
\begin{align}
{\nu}_{i0,\gamma} = [\gamma {\nu}_{i0,{\rho_s}}^{-1} +(1-\gamma){\nu}^{-1}_{i0,{\rho_{s+1}}}]^{-1},
\end{align}
and for nodes $i$ with $\bar d_i=1$, define
\begin{align}
 {\mu}_{0i,\gamma} = [\gamma {\mu}_{0i,{\rho_s}}^{-1} +(1-\gamma){\mu}^{-1}_{0i,{\rho_{s+1}}}]^{-1}.
\end{align}
\begin{lemma}
Given $({\nu}_{ij,{\rho_s}}, {\mu}_{ij,{\rho_s}}, Y_{\rho_s},|
{\nu}_{i0,{\rho_s}}, {\mu}_{0i,{\rho_s}})$ and $({\nu}_{ij,{\rho_{s+1}}},
{\mu}_{ij,{\rho_{s+1}}}, Y_{\rho_{s+1}},| {\nu}_{i0,{\rho_{s+1}}},
{\mu}_{0i,{\rho_{s+1}}})$, $j\in S^\varphi_i$, satisfying the
LMI (\ref{LMI Dircted uncertain}).
Then $({\nu}_{ij,{\gamma}}, {\mu}_{ij,{\gamma}}, Y_{\gamma},| {\nu}_{i0,{\gamma}}, {\mu}_{0i,{\gamma}})$ also satisfies the LMI (\ref{LMI Dircted uncertain}).

\end{lemma}

\emph{Proof: }
The statement of this lemma follows from the observation that
the inequality
(\ref{LMI Dircted uncertain}) is linear with respect to the variables $Y$
and $(\frac{1}{\nu_{ij,{\rho_s}}}, \frac{1}{\mu_{ij,{\rho_s}}}$ and
$\frac{1}{\nu_{i0,{\rho_s}}}$, $\frac{1}{\mu_{0i,{\rho_s}}})$ where these
latter variables appear.
\hfill$\Box$

Using the above lemma, we now define a collection of interpolated gains for
the control protocol (\ref{controller}), as follows. Suppose the collection
of positive constants $\beta_s$, $\beta_{s+1}$ and the grid points $\Gamma_\ell$ have the following properties: If $ \rho \in
[\rho_s, \rho_{s+1}]$, then
\begin{align}
(A(\rho)-A({\rho_s}))'(A(\rho)-A({\rho_s}))&\leq \beta_s^2 I, \quad \rho_s\leq \rho < \bar \rho_s,  \\
(A(\rho)-A({\rho_{s+1}}))'(A(\rho)-A({\rho_{s+1}}))&\leq \beta_{s+1}^2 I, \quad \underline{\rho}_{s+1}\leq \rho < \rho_{s+1},
\end{align}
where $\rho_s < \underline{\rho}_{s+1}< \bar \rho_s < \rho_{s+1}$. Define
$\gamma=\gamma(\rho)=\frac{\bar \rho_s- \rho}{\bar \rho_s -
  \underline{\rho}_{s+1}}$,
and let
\begin{align}
\label{definitionY}
Y_{\rho} &=\begin{cases}
 Y_{\rho_s}, ~\rho \in [\rho_s, \underline{\rho}_{s+1}),  \\
 Y_{\gamma},  ~\rho \in [\underline{\rho}_{s+1}, \bar \rho_s],   \\
 Y_{\rho_{s+1}},  ~\rho \in (\bar \rho_s, \rho_{s+1}],
\end{cases},\\
%\end{align}
%\begin{align}
{\nu}_{ij,\rho}& =\begin{cases}
{\nu}_{ij,\rho_s}, ~\rho \in [\rho_s, \underline{\rho}_{s+1}),  \\
{\nu}_{ij,\gamma},  ~\rho \in [\underline{\rho}_{s+1}, \bar \rho_s],   \\
{\nu}_{ij,\rho_{s+1}},  ~\rho \in (\bar \rho_s, \rho_{s+1}],
\end{cases},\\
%\end{align}
%\begin{align}
{\mu}_{ij,\rho}& =\begin{cases}
{\mu}_{ij,\rho_s}, ~\rho \in [\rho_s, \underline{\rho}_{s+1}),  \\
{\mu}_{ij,\gamma},  ~\rho \in [\underline{\rho}_{s+1}, \bar \rho_s],   \\
{\mu}_{ij,\rho_{s+1}},  ~\rho \in (\bar \rho_s, \rho_{s+1}],
\end{cases}.
\end{align}
Also, for nodes $i$ such that $d_i=1$, define
\begin{align}
\label{definitionY.d_i=0}
{\nu}_{i0,\rho}& =\begin{cases}
{\nu}_{i0,\rho_s}, ~\rho \in [\rho_s, \underline{\rho}_{s+1}),  \\
{\nu}_{i0,\gamma},  ~\rho \in [\underline{\rho}_{s+1}, \bar \rho_s],   \\
{\nu}_{i0,\rho_{s+1}},  ~\rho \in (\bar \rho_s, \rho_{s+1}].
\end{cases}
\end{align}
Likewise, for nodes $i$ such that $\bar d_i=1$, define
\begin{align}
{\mu}_{0i,\rho}& =\begin{cases}
{\mu}_{0i,\rho_s}, ~\rho \in [\rho_s, \underline{\rho}_{s+1}),  \\
{\mu}_{0i,\gamma},  ~\rho \in [\underline{\rho}_{s+1}, \bar \rho_s],   \\
{\mu}_{0i,\rho_{s+1}},  ~\rho \in (\bar \rho_s, \rho_{s+1}].
\end{cases}
\end{align}
Next, define the gain for the control protocol (\ref{controller})
\begin{align}
\label{final gain}
K_i(\rho)=- (\vartheta_i \sigma)^{-1}R^{-1}B_1'Y_{\rho}^{-1}.
\end{align}

The function $K_i$ is a continuous function on $\Gamma$, since $Y_{\rho}>0$ for all $\gamma \in [0, 1]$. Let $\Gamma_0$ be a set consisting of all the corner points $\underline{\rho}_{s+1}, \bar \rho_s \in \Gamma$. Without loss of generality, it is assumed that the set $\{ t \geq 0 \colon \rho(t) \in \Gamma_0 \}$ has zero Lebesgue measure.

The following theorem is the main result of this paper.
\begin{theorem}\label{Theorem 2}
Under Assumption \ref{graph assumption}, suppose that the time-varying parameter $\rho(\cdot)$ of the uncertain linear system (\ref{all agents dymamic}) satisfies the
condition
\begin{align}
\label{condition TH3}
\sup \limits_t|\dot \rho(t)| \le \frac{\eta}{\varrho}q, \quad \varrho
\triangleq \sup\limits_{\rho \in \Gamma \backslash \Gamma_0}\|\frac{d
  Y^{-1}_\rho}{d \rho}\|,
\end{align}
where $\eta= \min \limits_{\rho \in \Gamma} \lambda_{\min} \big (Y_\rho^{-1}B_1 R^{-1} B_1'Y_\rho^{-1}+ \bar Q \big)$ and $q \in  [0, 1)$ is a constant.

Then the control protocol (\ref{controller}) with the gain schedule
$K_i(\cdot)$ of the form (\ref{final gain}) solves the leader following tracking control Problem~\ref{prob 1} for the system (\ref{all agents dymamic}). Furthermore, this protocol guarantees the following
performance bound
 \begin{align} \label{cost function TH2}
 \sup_{\Xi}\mathcal {J}(u)  \le\frac{\hat \lambda}{(1-q)\sigma^2}\sum_{i=1}^{N}(x_0(0)-x_i(0))'(Y_{\rho(0)})^{-1}(x_0(0)-x_i(0)).
\end{align}
\end{theorem}

\emph{Proof: }
Since the matrix $Y^{-1}_\rho$ is continuous and piecewise differentiable except
at $\rho \in \Gamma_0$, it follows from \cite{Stilwell1999} that, given any $\varepsilon > 0$, there exists a continuous differentiable matrix function $X_\rho$ defined on $\Gamma$, and a constant $\varsigma>0$ for any corner points $\rho_c \in \Gamma_0$ such that
\begin{eqnarray}
&&\label{Smooth 1} \sup\limits_{\rho \in \Gamma}\|X_\rho -Y^{-1}_\rho\|< \varepsilon, \quad \rho \in (\rho_c - \varsigma, \rho_c + \varsigma),\\
&&\label{Smooth 1.1} X_\rho =Y^{-1}_\rho, \quad \rho \notin (\rho_c - \varsigma, \rho_c + \varsigma),\\
&&\label{Smooth 2} \sup\limits_{\rho \in \Gamma}\|\frac{d X_\rho}{d \rho}\|<  \sup\limits_{\rho \in \Gamma \backslash \Gamma_0}\|\frac{d Y^{-1}_\rho}{d \rho}\| =  \max \limits_{s=1,\ldots, M}\sup\limits_{\rho \in [\underline{\rho}_{s+1}, \bar \rho_s]} \|\frac{d Y^{-1}_\rho}{d \rho}\| \nonumber\\
&&= \max \limits_{s=1,\ldots, M}\max \limits_{\gamma \in [0, 1]} \|Y_{\gamma}^{-1}(Y_{\rho_s}-Y_{\rho_{s+1}})Y_{\gamma}^{-1}\|.
\end{eqnarray}
Note that the approximating matrix $X_\rho$ can be chosen symmetric, since $Y^{-1}_\rho$ is symmetric. Also by selecting a sufficiently small $\varepsilon>0$, a positive definite matrix $X_\rho$ can be selected for all $\rho \in \Gamma$.

Consider the closed loop large-scale
interconnected system describing tracking error dynamics of the system
(\ref{agents dymamic}) and (\ref{leader dymamic})
\begin{eqnarray} \label{error dymamic}
\dot{e}_i & =&A(\rho(t))e_i + B_1K_i(\rho)(\sum \limits_{j\in S^c_i}(e_i-e_j) +g_ie_i) \nonumber \\
&-&B_2 \sum\limits_{j\in S^\varphi_i}\varphi_{ij} (t, e_i-e_j)  - B_2 \sum\limits_{k\colon \bar d_k=1} \varphi_{0k}(t,e_k) - B_2 d_i \varphi_{i0}(t,e_i).
\end{eqnarray}
Let the following Lyapunov function candidate for this system be chosen
\begin{align}
V(e)=\sum_{i=1}^{N} e_i'X_{\rho}e_i.
\end{align}
We have

\begin{align}
\label{lyapunov equation11}
 \frac{d V(e)}{dt}=& \sum \limits_{i=1}^N 2 e_i' X_\rho \Big(A(\rho)e_i + B_1K_i(\rho)(\sum \limits_{j\in S^c_i}(e_i-e_j) +g_ie_i) \Big)  \nonumber\\
& - 2\sum\limits_{i=1}^N \sum\limits_{k\colon \bar d_k=1} e_i' X_\rho B_2\varphi_{0k}(t,e_k) + 2\sum \limits_{i=1}^N \sum\limits_{j\in S^\varphi_i}e_i' X_\rho B_2  \varphi_{ij} (t, e_j)\nonumber\\
& - 2\sum \limits_{i=1}^N \sum\limits_{j\in S^\varphi_i} e_i' X_\rho B_2  \varphi_{ij} (t, e_i) - 2\sum\limits_{i=1}^Nd_i e_i' X_\rho B_2 \varphi_{i0}(t,e_i) + \sum_{i=1}^{N}e_i' \dot X_\rho e_i.
\end{align}

Next, a bound on $\sum \limits_{i=1}^N 2e_i'X_\rho B_1K_i(\rho)(\sum
\limits_{j\in S^c_i}(e_i-e_j)+g_ie_i)$ is obtained. First we transform this expression
\begin{align}
\label{one trick}
&\sum \limits_{i=1}^N 2e_i'X_\rho B_1K_i(\rho)(\sum \limits_{j\in S^c_i}(e_i-e_j)+g_ie_i) \nonumber\\
&=-2e'(\Theta(\mathcal {L}^c_0+G)\otimes (X_\rho B_1(\sigma R)^{-1}B_1'Y_{\rho}^{-1}))e \nonumber\\
&=-2e'(\Theta(\mathcal {L}^c_0+G)\otimes (X_\rho B_1(\sigma R)^{-1}B_1'X_\rho))e \nonumber\\
&+2e'\Big(\Theta(\mathcal {L}^c_0+G)\otimes \big(X_\rho B_1(\sigma R)^{-1}B_1'(X_\rho-Y_{\rho}^{-1})\big)\Big)e.
\end{align}
Let $J=(\sigma R)^{-1/2}B_1'$ and consider
\begin{align}
\label{square}
&\Big ( \big(\sqrt \varepsilon (\mathcal {L}^c_0+G)'\Theta \big )\otimes J X_\rho -\frac{1}{\sqrt \varepsilon} I_N \otimes J(X_\rho-Y_{\rho}^{-1})\Big)' \nonumber\\
&\times \Big( \big(\sqrt \varepsilon (\mathcal {L}^c_0+G)'\Theta \big )\otimes J X_\rho -\frac{1}{\sqrt \varepsilon} I_N \otimes J(X_\rho-Y_{\rho}^{-1})\Big) \nonumber\\
&= \varepsilon \big(\Theta(\mathcal {L}^c_0+G)(\mathcal {L}^c_0+G)'\Theta \big) \otimes X_\rho J'J X_\rho  + \frac{1}{\varepsilon} I_N \otimes (X_\rho-Y_{\rho}^{-1}) J'J (X_\rho-Y_{\rho}^{-1}) \nonumber\\
&- \big(\Theta(\mathcal {L}^c_0+G) \big) \otimes X_\rho J'J (X_\rho-Y_{\rho}^{-1}) -\big((\mathcal {L}^c_0+G)'\Theta \big) \otimes(X_\rho-Y_{\rho}^{-1}) J'J X_\rho \ge 0.
\end{align}

It follows from (\ref{one trick}) and (\ref{square}) that
\begin{align}
\label{one trick 1}
& \sum\limits_{i=1}^N 2e_i'X_\rho B_1K_i(\rho)(\sum \limits_{j\in
  S^c_i}(e_i-e_j)+g_ie_i) \nonumber\\
&\leq -2e'(\Theta(\mathcal {L}^c_0+G)\otimes (X_\rho J'J X_{\rho}))e \nonumber\\
&+ e'\Big( \varepsilon \big(\Theta(\mathcal {L}^c_0+G)(\mathcal {L}^c_0+G)'\Theta \big) \otimes X_\rho J'J X_\rho + \frac{1}{\varepsilon} I_N \otimes (X_\rho-Y_{\rho}^{-1}) J'J (X_\rho-Y_{\rho}^{-1})\Big)e \nonumber\\
&= -\bar y'\big(H \otimes I_p\big)\bar y  + e'\Big( \varepsilon \big(\Theta(\mathcal {L}^c_0+G)(\mathcal {L}^c_0+G)'\Theta \big) \otimes X_\rho J'J X_\rho  \nonumber\\
&+ \frac{1}{\varepsilon} I_N \otimes (X_\rho-Y_{\rho}^{-1}) J'J (X_\rho-Y_{\rho}^{-1})\Big)e \nonumber\\
&\leq -2 \sigma \bar y'\big(I_N\otimes I_p\big)\bar y + \sum \limits_{i=1}^N \varepsilon\zeta \|JX_{\rho}\|^2\|e_i\|^2 + \sum \limits_{i=1}^N \varepsilon \|J\|^2\|e_i\|^2 \nonumber\\
&= -2\sigma e'\big(I_N \otimes X_\rho B_1 (\sigma R)^{-1}B_1' X_\rho \big)e + \sum \limits_{i=1}^N \varepsilon\zeta \|JX_{\rho}\|^2\|e_i\|^2  + \sum \limits_{i=1}^N \varepsilon \|J\|^2\|e_i\|^2 \nonumber\\
&=-2\sum \limits_{i=1}^N  e_i' X_\rho B_1 R ^{-1}B_1'X_\rho e_i + \sum \limits_{i=1}^N \varepsilon\zeta \|JX_{\rho}\|^2\|e_i\|^2  + \sum \limits_{i=1}^N \varepsilon \|J\|^2\|e_i\|^2,
\end{align}
where $\bar y=(I_N \otimes (\sigma R)^{-1/2}B_1'X_\rho) e$ and $\zeta=\lambda_{\min}\big(\Theta(\mathcal {L}^c_0+G)(\mathcal {L}^c_0+G)'\Theta\big)$.

Substituting (\ref{one trick 1}) into (\ref{lyapunov equation11}), we have
\begin{align}
\label{lyapunov equation112}
\frac{d V(e)}{dt}\leq & \sum \limits_{i=1}^N 2 e_i' X_\rho\Big( A(\rho) - B_1R^{-1}B_1'X_\rho \Big)e_i + 2\sum \limits_{i=1}^N \sum\limits_{j\in S^\varphi_i}e_i' X_\rho B_2  \varphi_{ij} (t, e_j) \nonumber \\
&  - 2\sum\limits_{i=1}^N \sum\limits_{k\colon \bar d_k=1} e_i' X_\rho B_2\varphi_{0k}(t,e_k) - 2\sum\limits_{i=1}^Nd_i e_i' X_\rho B_2 \varphi_{i0}(t,e_i)  \nonumber \\
& - 2\sum \limits_{i=1}^N \sum\limits_{j\in S^\varphi_i} e_i' X_\rho B_2  \varphi_{ij} (t, e_i) + \sum_{i=1}^{N}e_i' \dot X_\rho e_i  \nonumber \\
&+ \sum \limits_{i=1}^N \varepsilon\zeta \|JX_{\rho}\|^2\|e_i\|^2  + \sum \limits_{i=1}^N \varepsilon \|J\|^2\|e_i\|^2.
\end{align}
Consider the expression $\sum \limits_{i=1}^N 2 e_i' X_\rho\Big( A(\rho)- B_1 R ^{-1}B_1'X_\rho \Big)e_i$
\begin{align}
\label{le1}
& \sum \limits_{i=1}^N 2 e_i' X_\rho\Big( A(\rho) -  B_1 R ^{-1}B_1'X_\rho \Big)e_i \nonumber\\
&= \sum \limits_{i=1}^N 2 e_i' X_\rho\Big( A(\rho) -  B_1 R ^{-1}B_1'(X_\rho - Y_{\rho}^{-1} +Y_{\rho}^{-1})\Big)e_i \nonumber\\
&\leq  \sum \limits_{i=1}^N 2 e_i' X_\rho\Big( A(\rho) - B_1 R^{-1}B_1' Y_{\rho}^{-1} \Big)e_i + \sum \limits_{i=1}^N \varepsilon\sigma e'_iX_\rho J'J X_\rho e_i \nonumber\\
& + \sum \limits_{i=1}^N \frac{1}{\varepsilon}\sigma e'_i(X_\rho - Y_{\rho}^{-1}) J'J(X_\rho - Y_{\rho}^{-1})  e_i \nonumber\\
& \le \sum \limits_{i=1}^N 2 e_i' X_\rho\Big( A(\rho) + \sigma \vartheta_i B_1K_i(\rho)\Big)e_i  + \sum \limits_{i=1}^N \varepsilon\sigma \|J X_{\rho}\|^2\|e_i\|^2  + \sum \limits_{i=1}^N \varepsilon\sigma \|J\|^2\|e_i\|^2.
\end{align}
Substituting (\ref{le1}) into (\ref{lyapunov equation112}), we obtain
\begin{align}
\label{le2}
\frac{d V(e)}{dt}\leq & \sum \limits_{i=1}^N 2 e_i' X_\rho\Big( A(\rho) + \sigma \vartheta_i B_1K_i(\rho) \Big)e_i + 2\sum \limits_{i=1}^N \sum\limits_{j\in S^\varphi_i}e_i' X_\rho B_2  \varphi_{ij} (t, e_j)\nonumber \\
& - 2\sum\limits_{i=1}^N \sum\limits_{k\colon \bar d_k=1} e_i' X_\rho B_2\varphi_{0k}(t,e_k) - 2 \sum\limits_{i=1}^N d_i e_i' X_\rho B_2 \varphi_{i0}(t,e_i) + \sum_{i=1}^{N}e_i' \dot X_\rho e_i \\
& - 2\sum \limits_{i=1}^N \sum\limits_{j\in S^\varphi_i} e_i' X_\rho B_2  \varphi_{ij} (t, e_i) + \sum \limits_{i=1}^N \varepsilon(\zeta +\sigma) \|JX_\rho\|^2\|e_i\|^2 + \sum \limits_{i=1}^N \varepsilon (1+\sigma)  \|J\|^2\|e_i\|^2.\nonumber
\end{align}

Next, we turn our attention to the LMIs (\ref{LMI Dircted
  uncertain}). Using the Schur complement, each LMI (\ref{LMI Dircted
  uncertain}) can be transformed into the following Riccati inequality
\begin{align}
\label{riccati inequality with variation1}
&A(\rho)Y_\rho+Y_\rho A(\rho)' -  B_1 R^{-1} B_1'  + \big(\sum \limits_{j\in S^\varphi_i}(\frac{1}{\nu_{ij,\rho}}+ \frac{1}{\mu_{ij,\rho}}) + \sum\limits_{k\colon \bar d_k=1} \frac{1}{\mu_{0k,\rho}} \big)B_2B_2'\nonumber \\
& + Y_{\rho}(\bar Q + \sum\limits_{j\colon i \in S^\varphi_j}\mu_{ji,\rho}C'_{ji}C_{ji} + \sum \limits_{j\in S^\varphi_i} \nu_{ij,\rho}C'_{ij}C_{ij})Y_{\rho}\nonumber \\
& + \left(\frac{1}{\nu_{i0,\rho}} B_2B_2' +
  \nu_{i0,\rho}Y_{\rho}C'_{i0}C_{i0}Y_{\rho}\right)   \nonumber \\
& \qquad\qquad\qquad \mbox{(this term is present only if $d_i=1$)} \nonumber \\
& + N \mu_{0i,\rho}Y_{\rho}C'_{0i}C_{0i}Y_{\rho} < 0 \\
& \qquad\qquad\qquad \mbox{(this term is present only if $\bar d_i=1$)}. \nonumber
\end{align}
Note that the last two terms only appear in the Riccati inequality
(\ref{riccati inequality with variation1}) for those nodes $i$ for which
$d_i=1$ and/or $\bar d_i=1$.
After pre- and post-multiplying (\ref{riccati inequality with variation1}) by $Y_{\rho}^{-1}$ and substituting (\ref{final gain}) into it, we obtain
\begin{align}
\label{riccati in2 with variation1}
&Y_\rho^{-1} \big(A(\rho) +\vartheta_i \sigma B_1 K_i(\rho)\big) + \big(A(\rho) +\vartheta_i \sigma B_1 K_i(\rho)\big)' Y_\rho^{-1} +  Y_\rho^{-1}B_1 R^{-1} B_1'Y_\rho^{-1}+ \bar Q   \nonumber \\
& + \big(\sum \limits_{j\in S^\varphi_i}(\frac{1}{\nu_{ij,\rho}}+ \frac{1}{\mu_{ij,\rho}})  +  \sum\limits_{k\colon \bar d_k=1} \frac{1}{\mu_{0k,\rho}} \big)Y_\rho^{-1}B_2B_2'Y_\rho^{-1} + \sum \limits_{j\in S^\varphi_i} \nu_{ij,\rho}C'_{ij}C_{ij} + \sum\limits_{j\colon i \in S^\varphi_j}\mu_{ji,\rho}C'_{ji}C_{ji} \nonumber \\
& + \left(\frac{1}{\nu_{i0,\rho}}Y_\rho^{-1}B_2B_2'Y_\rho^{-1} +
  \nu_{i0,\rho}C'_{i0}C_{i0}\right) \nonumber \\
& \qquad\qquad\qquad \mbox{(this term is present only if $d_i=1$)} \nonumber \\
& + N \mu_{0i,\rho}C'_{0i}C_{0i} < 0 \\
& \qquad\qquad\qquad \mbox{(this term is present only if $\bar d_i=1$)}. \nonumber
\end{align}

Since the set $\Gamma$ is compact and coefficients of the
Riccati inequality (\ref{riccati in2 with variation1}) are continuous in
$\rho$, then provided $\varepsilon$ in (\ref{Smooth 1}) is sufficiently
small, replacing $Y^{-1}_\rho$ with $X_\rho$ in (\ref{riccati in2 with
  variation1}), except the term $Y_\rho^{-1}B_1 R^{-1} B_1'Y_\rho^{-1}$,
preserves the strict inequality:
\begin{align}
\label{riccati in2 with variation2}
&X_\rho\big(A(\rho) +\vartheta_i \sigma B_1 K_i(\rho)\big) + \big(A(\rho) +\vartheta_i \sigma B_1 K_i(\rho)\big)' X_\rho +  Y_\rho^{-1}B_1 R^{-1} B_1'Y_\rho^{-1}+ \bar Q   \nonumber \\
& + \big(\sum \limits_{j\in S^\varphi_i}(\frac{1}{\nu_{ij,\rho}}+ \frac{1}{\mu_{ij,\rho}})  +  \sum\limits_{k\colon \bar d_k=1} \frac{1}{\mu_{0k,\rho}} \big)X_\rho B_2B_2'X_\rho + \sum \limits_{j\in S^\varphi_i} \nu_{ij,\rho}C'_{ij}C_{ij} + \sum\limits_{j\colon i \in S^\varphi_j}\mu_{ji,\rho}C'_{ji}C_{ji} \nonumber \\
& + \left(\frac{1}{\nu_{i0,\rho}}X_\rho B_2B_2'X_\rho +
  \nu_{i0,\rho}C'_{i0}C_{i0}\right) \nonumber \\
& \qquad\qquad\qquad \mbox{(this term is present only if $d_i=1$)} \nonumber \\
& + N \mu_{0i,\rho}C'_{0i}C_{0i}< 0 \\
& \qquad\qquad\qquad \mbox{(this term is present only if $\bar d_i=1$)}. \nonumber
\end{align}

Using the Riccati inequality (\ref{riccati in2 with variation2}), and the following identities
\begin{align*}
\sum\limits_{i=1}^{N} \sum\limits_{j\in S^\varphi_i} \mu_{ij,\rho}e'_j C'_{ij}C_{ij}e_j =& \sum\limits_{i=1}^{N} \sum\limits_{j\colon i \in S^\varphi_j} \mu_{ji,\rho}e'_i C'_{ji}C_{ji}e_i,  \\
N \sum \limits_{i\colon \bar d_i=1}^N e_i' \mu_{0i,\rho}C'_{0i}C_{0i}e_i '=&N \sum \limits_{k\colon \bar d_k=1}^N e_k' \mu_{0k,\rho}C'_{0k}C_{0k}e_k, \\
\sum\limits_{i=1}^N d_i e_i' X_\rho B_2 \varphi_{i0}(t,e_i)=&\sum\limits_{i\colon d_i=1} e_i' X_\rho B_2 \varphi_{i0}(t,e_i),
\end{align*}
and completing the squares, it follows from (\ref{le2}) that
\begin{align}
&\int_{0}^{t}\frac{d V(e)}{dt}dt \leq -\sum_{i=1}^{N}\int_{0}^{t}e_i' \Big( Y_\rho^{-1}B_1 R^{-1} B_1'Y_\rho^{-1} + \bar Q \Big)e_i dt + \sum_{i=1}^{N}\int_{0}^{t} e_i' (\dot X_\rho)e_i dt \nonumber \\
&+ \sum \limits_{i=1}^{N}\sum\limits_{j\in S^\varphi_i} \nu_{ij,\rho} (\|\varphi_{ij} (t, e_i)\|^2-\|C_{ij}e_i\|^2 ) + \sum \limits_{i=1}^{N}\sum\limits_{j\in S^\varphi_i} \mu_{ij,\rho}(\| \varphi_{ij} (t, e_j)\|^2 - \|C_{ij}e_j\|^2) \nonumber \\
&+  \sum\limits_{i\colon d_i=1}  \nu_{i0,\rho} (\|\varphi_{i0} (t, e_i)\|^2-\|C_{i0}e_i\|^2 ) + N  \sum\limits_{k\colon \bar d_k=1}  \mu_{0k,\rho} (\|\varphi_{0k} (t, e_k)\|^2-\|C_{0k}e_k\|^2 ) \nonumber \\
&+ \sum \limits_{i=1}^N \int_{0}^{t}\varepsilon\big((\zeta +\sigma) \|JX_\rho\|^2 +(1+\sigma)  \|J\|^2 \big)\|e_i\|^2 dt.
\end{align}
Furthermore, using the norm-bounded condition (\ref{nb}), we obtain
\begin{align}
&\int_{0}^{t}\Big( \frac{d V(e)}{dt} + \sum_{i=1}^{N}e_i' (Y_\rho^{-1}B_1 R^{-1} B_1'Y_\rho^{-1}+ \bar Q)e_i \Big) dt\nonumber\\
& \leq \sum \limits_{i=1}^N \int_{0}^{t}\varepsilon\big((\zeta +\sigma) \|JX_\rho\|^2 +(1+\sigma)  \|J\|^2 \big)\|e_i\|^2 dt  + \sum_{i=1}^{N}\int_{0}^{t} e_i' (\dot X_\rho)e_i dt.
\end{align}
It follows from (\ref{condition TH3}) that $\dot Y_{\rho}^{-1}$ satisfies
the condition
\begin{align}
\|\dot Y_{\rho}^{-1}\| \leq q \lambda_{\min} (Y_\rho^{-1}B_1 R^{-1} B_1'Y_\rho^{-1}+ \bar Q ).
\end{align}
Together with (\ref{Smooth 2}), this inequality yields
\begin{align}
\|\dot X_\rho\| &\leq q \lambda_{\min}(Y_\rho^{-1}B_1 R^{-1} B_1'Y_\rho^{-1}+ \bar Q).
\end{align}
Since
\begin{align*}
\lambda_{\min}(Y_\rho^{-1}B_1 R^{-1} B_1'Y_\rho^{-1}+ \bar Q)I \leq Y_\rho^{-1}B_1 R^{-1} B_1'Y_\rho^{-1}+ \bar Q,
\end{align*}
we have
\begin{align}
\label{V1-V1}
  V(e(t)) - V(e(0)) \leq& - \sum_{i=1}^{N}\int_{0}^{t}e_i' \Big( Y_\rho^{-1}B_1 R^{-1} B_1'Y_\rho^{-1}  + \bar Q \Big)e_i dt \nonumber \\
 & +q \sum_{i=1}^{N}\int_{0}^{t} e_i' (Y_\rho^{-1}B_1 R^{-1} B_1'Y_\rho^{-1}+ \bar Q)e_i dt \nonumber \\
 &+\sum \limits_{i=1}^N \int_{0}^{t}\varepsilon\big((\zeta +\sigma) \|JX_\rho\|^2 +(1+\sigma)  \|J\|^2 \big)\|e_i\|^2 dt .
\end{align}
Since $V(e(t)) \ge 0$, (\ref{V1-V1}) implies
\begin{align}
\label{V1-V1.1}
& (1-q)\sum_{i=1}^{N}\int_{0}^{t}e_i'( Y_\rho^{-1}B_1 R^{-1} B_1'Y_\rho^{-1}+ \bar Q)e_i dt \nonumber \\
& - \sum \limits_{i=1}^N \int_{0}^{t}\varepsilon\big((\zeta +\sigma) \|JX_\rho\|^2 +(1+\sigma) \|J\|^2 \big)\|e_i\|^2 dt < V(e(0)).
\end{align}
We now choose $\varepsilon>0$ to be sufficiently
small to ensure that
\begin{eqnarray*}
\varepsilon_1 &\triangleq& (1-q)\eta -\varepsilon \left((\zeta +\sigma)
(\max_{\rho\in\Gamma} \|Y_\rho^{-1}\| +\varepsilon)^2 +(1+\sigma)\right)
\|J\|^2
\end{eqnarray*}
is positive. Such an $\varepsilon>0$ exists
since at $\varepsilon=0$, $\varepsilon_1=(1-q)\eta>0$, and as function of
$\varepsilon$, $\varepsilon_1$ is continuous at $\varepsilon=0$. Then with
this $\varepsilon$,
\begin{eqnarray*}
0<\varepsilon_1 < (1-q)\eta
-\varepsilon \left((\zeta +\sigma)
\|JX_\rho\|^2 +(1+\sigma)\|J\|^2\right),
\end{eqnarray*}
and it follows from (\ref{V1-V1.1}) that
\begin{align*}
 \sum_{i=1}^{N}\int_{0}^t \|e_i\|^2dt\le \frac{1}{\varepsilon_1}\sum_{i=1}^{N}\big( \|(Y_{\rho(0)})^{-1}\| + \varepsilon \big)\|e_i(0)\|^2.
\end{align*}
The above inequality holds for all $t>0$ and the right-hand side is
independent of $t$, therefore we conclude that
$\lim_{t\to\infty}\int_{0}^t \|e_i\|^2dt $ exists and is finite.
This allows us to let $t\to \infty$ in (\ref{V1-V1.1}) to obtain
\begin{align}
\label{cost-V1}
& (1-q)\sum_{i=1}^{N}\int_{0}^{\infty}e_i'(Y_\rho^{-1}B_1 R^{-1} B_1'Y_\rho^{-1}+ \bar Q)e_i dt \nonumber \\
& \le \sum \limits_{i=1}^N \int_{0}^{\infty}\varepsilon\big((\zeta +\sigma) \|JX_\rho\|^2 +(1+\sigma) \|J\|^2 \big)\|e_i\|^2 dt +\sum_{i=1}^{N}e_i(0)'X_{\rho(0)}e_i(0).
\end{align}

Note that the left hand side of (\ref{cost-V1}) is independent of
$\varepsilon$. Then we can let $\varepsilon \to 0$ in
(\ref{cost-V1}). Since $X_{\rho(0)} \rightarrow
(Y_{\rho(0)})^{-1}$ as $\varepsilon \to 0$, this leads to
\begin{align}
\sum_{i=1}^{N}\int_{0}^{\infty}e_i' (Y_\rho^{-1}B_1 R^{-1} B_1'Y_\rho^{-1}+ \bar Q)e_i dt \le \frac{1}{1-q} \sum_{i=1}^{N}e_i'(0)(Y_{\rho(0)})^{-1} e_i(0).
\end{align}
Using (\ref{cost function}) and (\ref{controller}), we have
\begin{align}
 \mathcal {J}(u) &\leq  \sum_{i=1}^{N}\int_{0}^{\infty}e_i' (\frac{\hat \lambda}{\sigma^2} Y_\rho^{-1}B_1 R^{-1} B_1'Y_\rho^{-1}+ Q)e_i dt \nonumber\\
 &=\frac{\hat \lambda}{\sigma^2} \sum_{i=1}^{N}\int_{0}^{\infty}e_i' ( Y_\rho^{-1}B_1 R^{-1} B_1'Y_\rho^{-1}+ \bar Q)e_i dt \nonumber\\
 &\leq  \frac{\hat \lambda}{(1-q)\sigma^2} \sum_{i=1}^{N}(x_0(0)-x_i(0))'(Y_{\rho(0)})^{-1}(x_0(0)-x_i(0)).
\end{align}

It implies that the control protocol (\ref{controller}) with $K_i(\cdot)$ of the form (\ref{final gain}) solves leader following tracking control Problem~\ref{prob 1} for the system (\ref{all agents dymamic}), and also guarantees
  the performance bound (\ref{cost function TH2}).
\hfill$\Box$

\begin{remark}
  It should be noted that the proposed solution depends on the global information on the system topology. Specifically, the constants
$\theta_i$ and $\sigma$ are determined by the communication topology (but do not
depend on the interconnection topology). On the other hand, the LMIs
(\ref{LMI Dircted}) are setup using the knowledge of interconnection topology.
\end{remark}

% \hfill $\Box$
\section{Example}
\label{example}
To illustrate the proposed method, consider a mass-spring-damper system in Fig.~\ref{blocks}. The system consists of $21$
identical masses which are coupled by different springs and dampers. Each
mass is also connected to the wall with a spring. The dynamics of the coupled system
are governed by the following equations

\begin{align}
\label{dynamic of MSD}
m\ddot{\alpha}_0=&k_{1,1}(\alpha_1 - \alpha_0) + k_{1,2}(\dot \alpha_1 - \dot \alpha_0) + k_{0,1}(\alpha_{20} - \alpha_0) + k_{0,2}(\dot \alpha_{20}- \dot \alpha_0)-\bar k \alpha_0, \nonumber\\
m\ddot{\alpha}_i=&k_{i,1}(\alpha_{i-1} - \alpha_i) + k_{i,2}(\dot \alpha_{i-1} - \dot \alpha_i) \nonumber \\
&+ k_{i+1,1}(\alpha_{i+1}-\alpha_i) + k_{i+1,2}(\dot \alpha_{i+1} - \dot \alpha_i) -\bar k \alpha_i + u_i, \quad i=1, \ldots, 19,  \\
m\ddot{\alpha}_{20}=&k_{20,1}(\alpha_{19} - \alpha_{20}) + k_{20,2}(\dot \alpha_{19} - \dot \alpha_{20})+ k_{0,1}(\alpha_0 - \alpha_{20})  + k_{0,2}(\dot \alpha_0 - \dot \alpha_{20}) -\bar k\alpha_{20} + u_{20}, \nonumber
\end{align}
where $\alpha_i$, $\dot \alpha_i$ are the displacement and the velocity of
mass $i$, $\bar k$ and $k_{i,1}$, $i=0, \ldots, 20$ are the spring
constants, $k_{i,2}$, $i=0, \ldots, 20$  are the damper coefficients, and
$m$ is the mass of each block.

Choosing the state vectors as $x_i=(\alpha_i, \dot{\alpha}_i), i=0, \ldots,
20$,  each equation in (\ref{dynamic of MSD}) can be written in the form of (\ref{agents dymamic}), (\ref{leader dymamic}),
where
\begin{eqnarray*}
&&A=\left[\begin{array}{cc}
                          0  &  1  \\
                        -\frac {\bar k}{m} & 0
                        \end{array}\right], \quad
                         B_1=\left[\begin{array}{c}
                          0    \\
                        \frac{1}{m}
                        \end{array}\right], \quad
                         B_2=\left[\begin{array}{c}
                                0   \\
                         \frac{1}{m}
                        \end{array}\right],\\
&& \varphi_{i,i-1}(z)=\varphi_{i-1,i}(z)=[k_{i,1}~k_{i,2}]z, \quad (i=1,\ldots, 20)\\
&& \varphi_{0,20}(z)=\varphi_{20,0}(z)=[k_{0,1}~k_{0,2}]z, \quad
z\in\mathbf{R}^2.
\end{eqnarray*}

To verify the norm-bounded condition, let us assume that
\[
[k_{i,1}~k_{i,2}]=\Delta_{i,i-1} C_{i,i-1}, \quad
[k_{0,1}~k_{0,2}]=\Delta_{0,20} C_{0,20},
\]
where $C_{i,i-1}=[k_{i,1}~ k_{i,2}]$, $C_{0,20}=[ k_{0,1}~
k_{0,2}]$, are given matrices, and $\Delta_{i,i-1}$, $\Delta_{0,20}\in
[0,1]$ are scalars. Then all the couplings between the agents satisfy the
norm bound condition (\ref{nb}). In this example, for simplicity we let
$ k_{i,1}= k_{i,2}= k$ for all $i=0,\ldots, 20$.

\begin{figure}
 \centering
 \scalebox{0.7}{
 \begin{tikzpicture}[->,>=stealth',shorten >=1pt,auto,node distance=2cm,
   thick,main node/.style={circle,fill=white!20,draw,font=\sffamily\bfseries},]

   \node[main node] (2) {~1~~};
   \node[main node] (3) [below of=2]{~2~~};
   \node[main node] (4) [right of=3] {~8~~};
   \node[main node] (5) [right of=4] {12};
   \node[main node] (6) [right of=5] {15};
   \node[main node] (7) [right of=6] {19};
   \node[main node] (8) [above of=7] {20};
   \node[main node] (1) [above of=5]{~0~~};

   \path[every node/.style={font=\sffamily\small}]
   (1) edge node {}  (2)
   (1) edge node {}  (4)
   (1) edge node {}  (5)
   (1) edge node {}  (6)
   (2) edge node {}  (3)
   (7) edge node {}  (8);
 \path[draw,dashed] (3)  -- (4);
 \path[draw,dashed] (4)  -- (5);
 \path[draw,dashed] (5)  -- (6);
 \path[draw,dashed] (6)  -- (7);

 \end{tikzpicture}
 }
 \centering
 \caption{Directed communication graph.}
 \label{control graph}
 \end{figure}
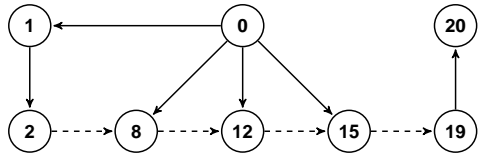

The structure of the system (\ref{dynamic of MSD}) suggests that the
agents are coupled according to the undirected cyclic graph shown in
Fig.~\ref{coupling graph}, since the force exerted by mass $i$ on mass
$i+1$ exactly reciprocates the force exerted by mass $i+1$ on mass $i$:
\begin{eqnarray*}
k_{i+1,1}(\alpha_{i+1}-\alpha_i) + k_{i+1,2}(\dot \alpha_{i+1} - \dot
\alpha_i) = - k_{i+1,1}(\alpha_{i} - \alpha_{i+1}) - k_{i+1,2}(\dot \alpha_{i} -
\dot \alpha_{i+1}).
\end{eqnarray*}
We will treat the graph in Fig.~\ref{coupling graph} as a special case of
directed graph with symmetric adjacency matrix.
On the other hand, the communication topology
of the system is assumed to be a directed graph shown
in Fig~\ref{control graph}. According to this graph, agents $1$,
$8$, $12$ and $15$ observe the leader.

 \begin{figure}

 \centering
 \scalebox{0.5}{
\begin{tikzpicture}[every node/.style={draw,outer sep=0pt,thick}]
\tikzstyle{spring}=[thick,decorate,decoration={zigzag,pre length=0.3cm,post
length=0.3cm,segment length=6}]
\tikzstyle{damper}=[thick,decoration={markings,
  mark connection node=dmp,
  mark=at position 0.5 with
  {
    \node (dmp) [thick,inner sep=0pt,transform shape,rotate=-90,minimum
width=15pt,minimum height=3pt,draw=none] {};
    \draw [thick] ($(dmp.north east)+(2pt,0)$) -- (dmp.south east) -- (dmp.south
west) -- ($(dmp.north west)+(2pt,0)$);
    \draw [thick] ($(dmp.north)+(0,-5pt)$) -- ($(dmp.north)+(0,5pt)$);
  }
}, decorate]
\tikzstyle{ground}=[fill,pattern=north east lines,draw=none,minimum
width=0.75cm,minimum height=0.3cm]

\node (M0) [minimum width=1.8cm, minimum height=1.8cm] {\LARGE$m_0$};
\node (M1) at (4,0) [minimum width=1.8cm, minimum height=1.8cm] {\LARGE$m_1$};

\node (M2) at (8,0)[minimum width=1.8cm, minimum height=1.8cm] {\LARGE$m_{19}$};
\node (M3) at (12,0) [minimum width=1.8cm, minimum height=1.8cm] {\LARGE$m_{20}$};

\draw[spring] ($(M0.east) + (0,0.6)$) -- ($(M1.west) + (0,0.6)$) node at (2,1.1)[draw=none, outer sep=2pt,name=xN-1] {\LARGE$k_{1,1}$};
\draw[damper] ($(M0.east) - (0,0.6)$) -- ($(M1.west) - (0,0.6)$) node at (2,-1.2)[draw=none, outer sep=2pt,name=xN-1] {\LARGE$k_{1,2}$};

\draw[spring] ($(M1.east) + (0,0.6)$) -- ($(M1.east) + (1.4,0.6)$); %node at (6,1.1)[draw=none, outer sep=2pt,name=xN-1] {\LARGE$k_{??1}$};
\draw[thick, dashed]($(M1.east) + (1.5,0.6)$)--($(M2.west) + (0,0.6)$);
\draw[damper] ($(M1.east) - (0,0.6)$) -- ($(M1.east) +(1.4,-0.6)$); %node at (6,-1.2) [draw=none, outer sep=2pt,name=xN-1] {\LARGE$k_{22}$};
\draw[thick, dashed]($(M1.east) + (1.5,-0.6)$)--($(M2.west) + (0,-0.6)$);

\draw[spring] ($(M2.east) + (0,0.6)$) -- ($(M3.west) + (0,0.6)$)node at (10,1.1)[draw=none, outer sep=2pt,name=xN-1] {\LARGE$k_{20,1}$};
\draw[damper] ($(M2.east) - (0,0.6)$) -- ($(M3.west) - (0,0.6)$)node at (10,-1.2)[draw=none, outer sep=2pt,name=xN-1] {\LARGE$k_{20,2}$};

\draw[spring] ($(M1.east) + (0,3.8)$) -- ($(M2.west) + (0,3.8)$)node at (6,4.2)[draw=none, outer sep=2pt,name=xN-1] {\LARGE$k_{0,1}$};
\draw[damper] ($(M1.east) + (0,2.6)$) -- ($(M2.west) + (0,2.6)$)node at (6,3.2)[draw=none, outer sep=2pt,name=xN-1] {\LARGE$k_{0,2}$};
\draw[thick] ($(M0.north) + (0, 3.2)$)-- ($(M0.north) + (0,0)$);
\draw[thick] ($(M3.north) + (0, 3.2)$)-- ($(M3.north) + (0,0)$);

\draw[thick] ($(M1.east) + (0,3.8)$)--($(M0.east) + (-0.9, 3.8)$);
\draw[thick] ($(M1.east) + (0,2.6)$)--($(M0.east) + (-0.9, 2.6)$);

\draw[thick] ($(M2.west) + (0,3.8)$)--($(M3.east) + (-0.9, 3.8)$);
\draw[thick] ($(M2.west) + (0,2.6)$)--($(M3.east) + (-0.9, 2.6)$);

\draw[thick, dashed] ($(M0.north east)$) -- ($(M0.north east) + (0,.65)$);
\draw[thick, dashed] ($(M1.north east)$) -- ($(M1.north east) + (0,.65)$);
\draw[thick, dashed] ($(M2.north east)$) -- ($(M2.north east) + (0,.65)$);
\draw[thick, dashed] ($(M3.north east)$) -- ($(M3.north east) + (0,.65)$);

\draw[ultra thick, -latex] ($(M0.north east) + (0,0.6)$) -- ($(M0.north east) + (0.6,0.6)$) node at (1.2,1.6) [draw=none,left,above, outer sep=2pt,name=xN-1] {\LARGE$\alpha_0$};
\draw[ultra thick, -latex] ($(M1.north east) + (0,0.6)$) -- ($(M1.north east) + (0.6,0.6)$)  node at (5.2,1.6) [draw=none,above,outer sep=2pt,name=xN-1] {\LARGE$\alpha_1$};
\draw[ultra thick, -latex] ($(M2.north east) + (0,0.6)$) -- ($(M2.north east) + (0.6,0.6)$)  node at (9.2,1.6) [draw=none,above,outer sep=2pt,name=xN-1] {\LARGE$\alpha_{19}$};
\draw[ultra thick, -latex] ($(M3.north east) + (0,0.6)$) -- ($(M3.north east) + (0.6,0.6)$)  node at (13.2,1.6) [draw=none,above,outer sep=2pt,name=xN-1] {\LARGE$\alpha_{20}$};

\draw[ultra thick, -latex] ($(M1.west) - (0.75,0)$) -- ($(M1.west)$)node at (2,0)[draw=none,outer sep=2pt,name=xN-1] {\LARGE$u_1$};
\draw[ultra thick, -latex] ($(M2.west) - (0.65,0)$) -- ($(M2.west)$)node at (6,0)[draw=none,outer sep=2pt,name=xN-1] {\LARGE$u_{19}$};
\draw[ultra thick, -latex] ($(M3.west) - (0.65,0)$) -- ($(M3.west)$)node at (10,0)[draw=none,outer sep=2pt,name=xN-1] {\Large$u_{20}$};

\draw[spring] ($(M0.east) + (-1.8,0)$) -- ($(M0.east) + (-3.8,0)$)node at (-1.8,0)[draw=none,above,outer sep=2pt,name=xN-1] {\LARGE$\bar k$};;

\draw[spring] ($(M1.east) + (-0.9,-2)$) -- ($(M1.east) + (-2.9,-2)$)node at (3,-2)[draw=none,above,outer sep=2pt,name=xN-1] {\LARGE$\bar k$};;
\draw[thick] ($(M1.east) +(-2.9,-2)$)  -- ($(M0.west)+ (-2,-2)$);
\draw[thick] ($(M1.east) + (-0.9,-2)$)-- ($(M1.south) + (0,0)$);

\draw[spring] ($(M2.east) + (-0.9,-3)$) -- ($(M2.east) + (-2.9,-3)$)node at (7,-3)[draw=none,above,outer sep=2pt,name=xN-1] {\LARGE$\bar k$};;
\draw[thick] ($(M2.east) +(-2.9,-3)$)  -- ($(M0.west)+ (-2,-3)$);
\draw[thick] ($(M2.east) + (-0.9,-3)$)-- ($(M2.south) + (0,0)$);

\draw[spring] ($(M3.east) + (-0.9,-4)$) -- ($(M3.east) + (-2.9,-4)$)node at (11,-4)[draw=none,above,outer sep=2pt,name=xN-1] {\LARGE$\bar k$};;
\draw[thick] ($(M3.east) +(-2.9,-4)$)  -- ($(M0.west)+ (-2,-4)$);
\draw[thick] ($(M3.east) + (-0.9,-4)$)-- ($(M3.south) + (0,0)$);

\node (wall) [ground, rotate=-90, minimum width=8.5cm,yshift=-3.05cm] {};
\draw (wall.north east) -- (wall.north west);
\end{tikzpicture}
}
\caption{Mass-spring-damper system.}
\label{blocks}
\end{figure}
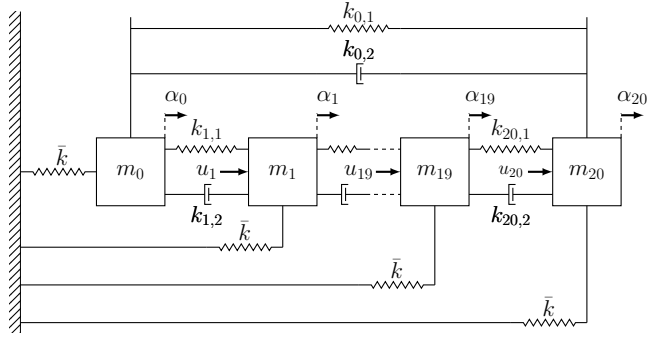

\begin{figure}
 \centering
 \scalebox{0.7}{
 \begin{tikzpicture}[->,>=stealth',shorten >=1pt,auto,node distance=2cm,
   %thick,main node/.style={circle,fill=blue!20,draw,font=\sffamily\Large\bfseries}]
   thick,main node/.style={circle,fill=white!20,draw,font=\sffamily\bfseries},]

  % \node[main node] (1) {0};
   \node[main node] (1)  {~0~~};
   \node[main node] (2) [above right of=1] {~1~~};
   \node[main node] (3) [ right of=2] {~2~~};
   \node[main node] (4) [ right of=3] {~3~~};
   \node[main node] (5) [below right of=1] {20};
   \node[main node] (6) [right of=5] {19};
   \node[main node] (7) [right of=6] {18};

   \path[every node/.style={font=\sffamily\small}]
     (1) %edge node [left] {0.6} (4)
        edge  (2)
        edge  (5)

     (2) edge  (1)
       edge (3)

     (3) edge  (2)
        edge  (4)%edge [bend right] node[right] {0.2} (3);
    (4) edge  (3)
        %edge (7)

    (7) %edge  (4)
        edge  (6)

    (6) edge (5)
        edge (7)

    (5) edge(1)
        edge  (6);
      \path[draw,dashed] (7)  -- (4);
      \path[draw,dashed] (4)  -- (7);
 \end{tikzpicture}
 }
 \centering
 \caption{Undirected coupling graph.}
 \label{coupling graph}
 \end{figure}
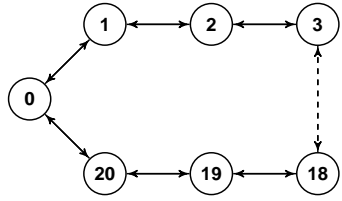

To illustrate the results of the paper, the protocol matrices were computed
using Theorem~\ref{Theorem 2}, and then
the trajectories of the coupled mass-spring-damper system with the obtained
protocol were simulated. To design the protocol, the parameters of the coupled
mass-spring-damper system were chosen to be $m=3$ kg, $k=0.1$ N/m, and $\bar
k=2.4-1.4\rho(t)$ N/m with $\rho(t)=\cos(t)$. The headway distance between
the mass bodies was assumed to be $1$ m.

In the simulation, the spring constants and damper coefficients were chosen
to be $k_{i,1}=0.1$ N/m and $k_{i,2}=0.1$ N/(m/s), $i=0, \ldots, 20$,
respectively. The initial states of the leader and the followers were set to the
values shown in table~\ref{initial states}. In the cost function, we chose
$Q=\begin{bmatrix}10 & 0\\0  & 100\end{bmatrix}$ and $R=0.001$.

\begin{table}[htbp]
\centering
    \caption{The initial states of the leader and followers.}  \label{initial states}
\newcommand{\tabincell}[2]{\begin{tabular}{@{}#1@{}}#2\end{tabular}}
  \centering
    \begin{tabular}{|c|c|
c|c|c|c|c|
c|c|c|c|}
 \hline
     \multicolumn{1}{|c|}{$x_0'(0)$} & \multicolumn{1}{|c|}{$x_1'(0)$} & \multicolumn{1}{|c|}{$x_2'(0)$} & \multicolumn{1}{|c|}{$x_3'(0)$} & \multicolumn{1}{|c|}{$x_4'(0)$}&\multicolumn{1}{|c|}{$x_5'(0)$} & \multicolumn{1}{|c|}{$x_6'(0)$} \\ \hline
[0.5 0]& [0.45 0] &  [0.4 0] & [0.3 0] & [0.2 0] &[0.15 0] &  [0.25 0]    \\ \hline
       \multicolumn{1}{|c|}{$x_7'(0)$} & \multicolumn{1}{|c|}{$x_8'(0)$} &\multicolumn{1}{|c|}{$x_9'(0)$}  & \multicolumn{1}{|c|}{$x_{10}'(0)$} & \multicolumn{1}{|c|}{$x_{11}'(0)$} & \multicolumn{1}{|c|}{$x_{12}'(0)$} & \multicolumn{1}{|c|}{$x_{13}'(0)$}  \\ \hline
    [0.35 0]   & [0.45 0] & [0.55 0] &  [0.65 0]   &  [0.55 0]   & [0.45 0] & [0.35 0]  \\ \hline
    \multicolumn{1}{|c|}{$x_{14}'(0)$} & \multicolumn{1}{|c|}{$x_{15}'(0)$} & \multicolumn{1}{|c|}{$x_{16}'(0)$} & \multicolumn{1}{|c|}{$x_{17}'(0)$} & \multicolumn{1}{|c|}{$x_{18}'(0)$} & \multicolumn{1}{|c|}{$x_{19}'(0)$} & \multicolumn{1}{|c|}{$x_{20}'(0)$} \\ \hline
    [0.45 0]   &  [0.5 0]   & [0.4 0] & [0.3 0.1] &  [0.2 0.2]   &  [0.1 0.3]   & [0 0.4]   \\ \hline
    \end{tabular}
\end{table}

To design the synchronization protocol, we chose $\Gamma=[-1, 1]$ with $4$
design points
\begin{align*}
\label{design points}
\Gamma_\ell= \big \{-1, -0.3333, 0.3333,  1 \big \}.
\end{align*}
Choosing $\beta_s=0.3111$ so that the properties (\ref{designpoints}) hold
for each design point, we solved the corresponding LMIs (\ref{LMI
  Dircted}). Next, we constructed a continuous-gain control protocol by
using the interpolation technique based on Theorem~\ref{Theorem 2}. We
computed $\eta=0.0143$, $\varrho=0.0082 $ and we need to
select $q\in (0, 1)$ such that $\sup_t|\dot \rho(t)| \le 1.7485q$,
$q \in (0, 1)$. Since
$\sup_t|\dot \rho(t)| = \sup_t|-\sin t|= 1$, this allows
us to choose $q =0.5750$. Then the theoretically predicted bound on the
performance (\ref{cost function  TH2}) is computed to be $ 329.1316$. On the
other hand, by simulating the
system on the time interval $[0, 12]$, we numerically found the value of
the performance cost (\ref{cost function}) for this particular instance of
the system to be equal to $\mathcal{J}(u)= 39.7982$.

\begin{figure}[htbp]
\centering
 \scalebox{0.8}{
\includegraphics[width=.95\columnwidth]{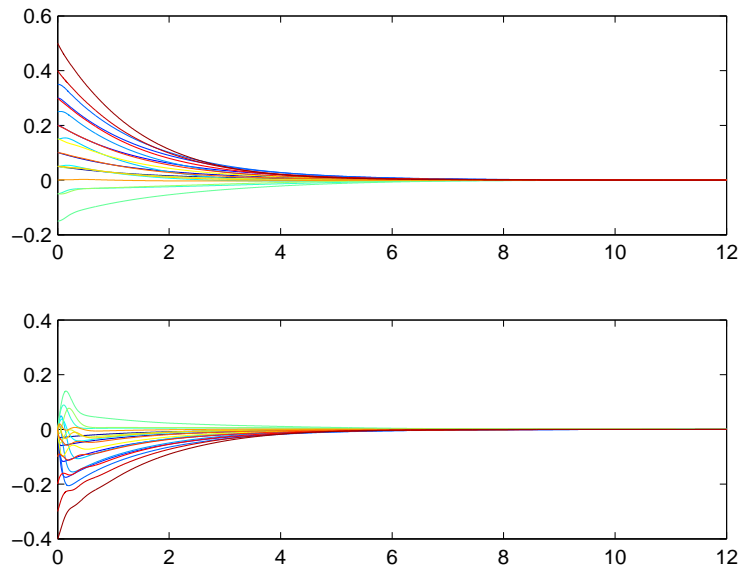}
}
\caption{Relative displacement (the top figure) and relative velocities of the
  followers with respect to the leader.}
\label{trackingerror}
\end{figure}

\begin{figure}[htbp]
\centering
 \scalebox{0.8}{
\includegraphics[width=.95\columnwidth]{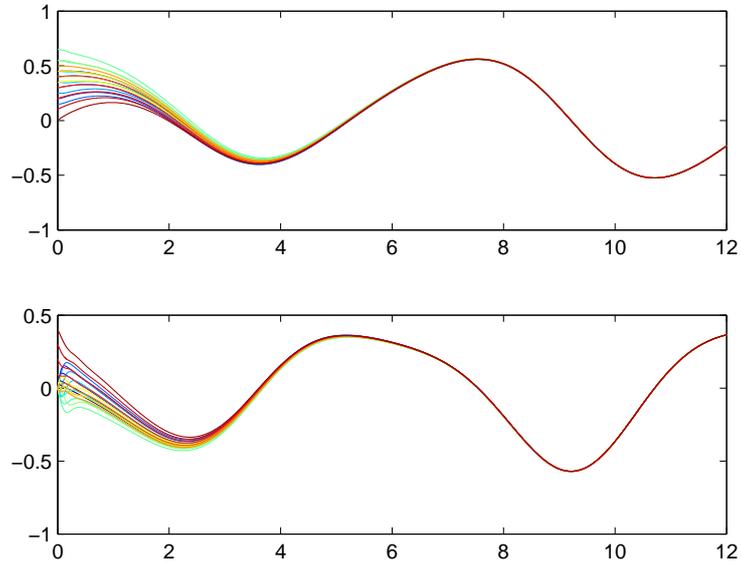}
}
\caption{%\textbf{VO: Please indicate which figure is velocity, and which one is
%   displacement. Then change the caption as follows:}
The displacement (the top figure) and velocity trajectories of the
  leader and followers.}
\label{trajectory}
\end{figure}

\begin{figure}[htbp]
\centering
 \scalebox{0.8}{
\includegraphics[width=.95\columnwidth]{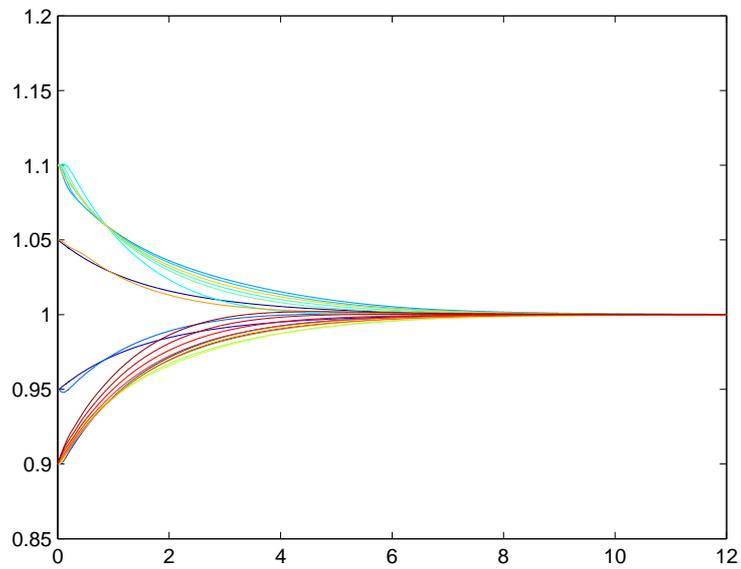}
}
\caption{The relative distance between the blocks.}
\label{distance}
\end{figure}

\begin{figure}[htbp]
  \centering
    \includegraphics[width=0.95\linewidth, height = 0.3\textheight]{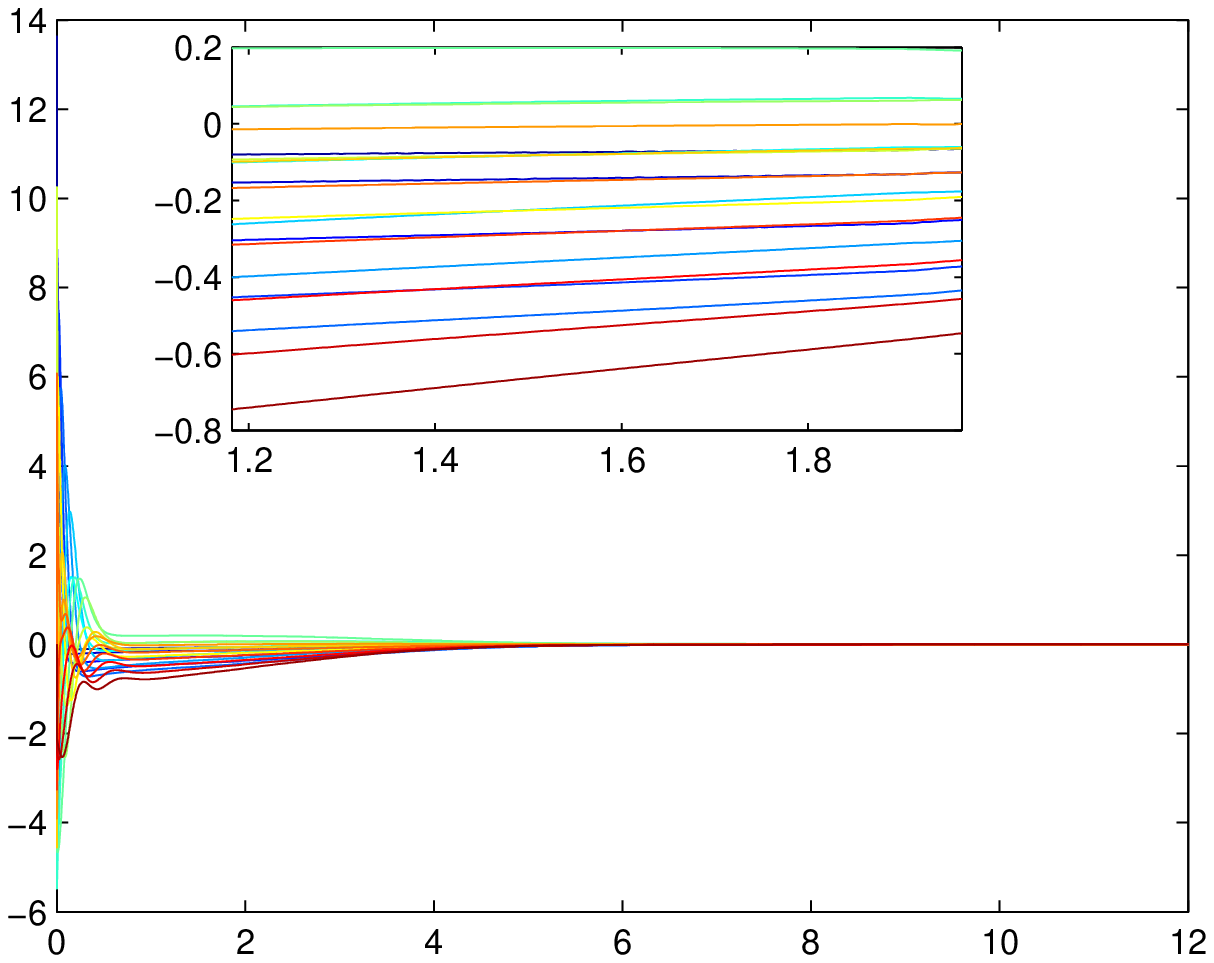} \\
     \includegraphics[width=0.95\linewidth, height = 0.3\textheight]{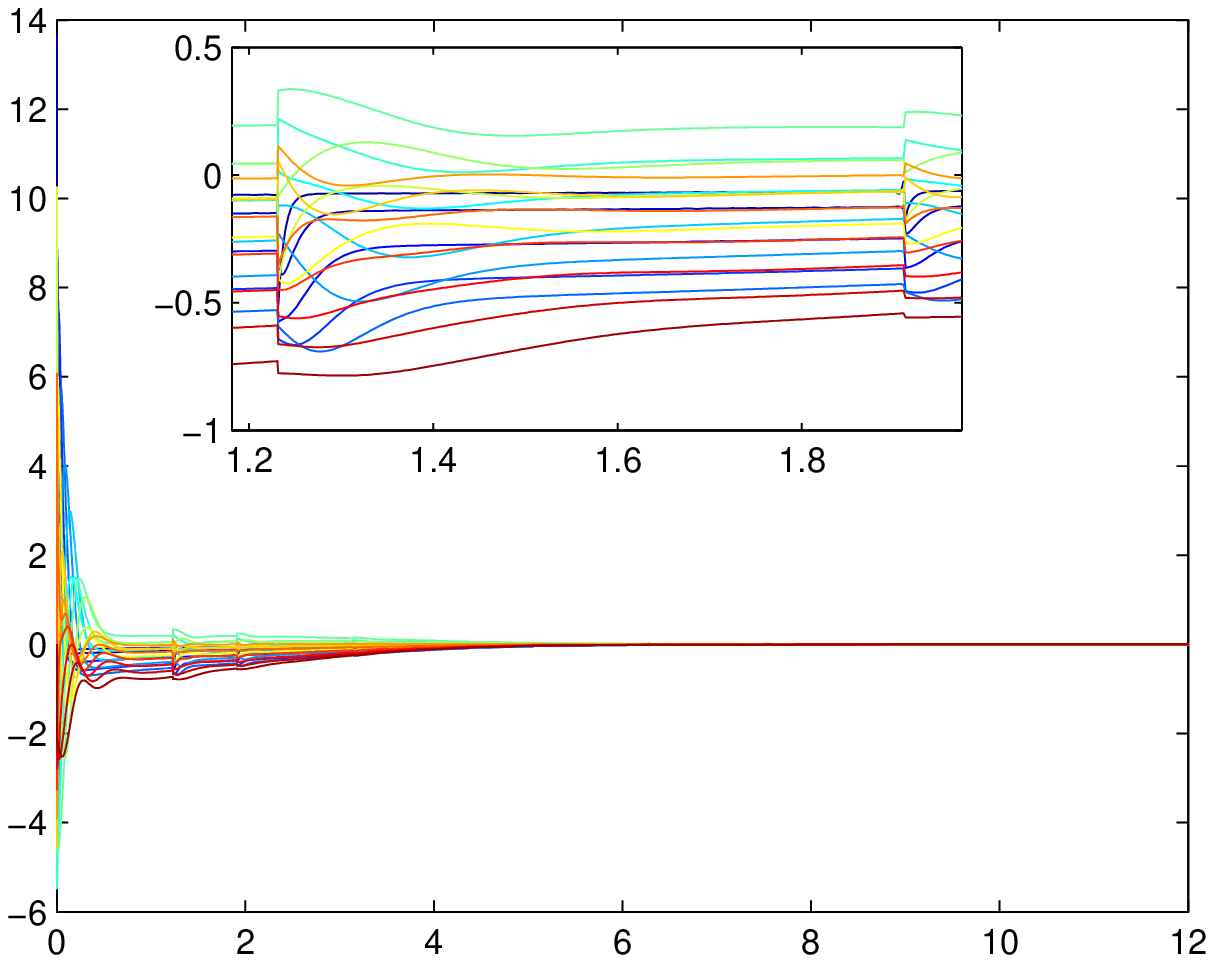}
\caption{The continuous control signals (the top figure) and switching control signals of the followers.}
\label{control_signals}
\end{figure}

\begin{figure}[htbp]
  \centering
    \includegraphics[width=0.95\linewidth, height = 0.3\textheight]{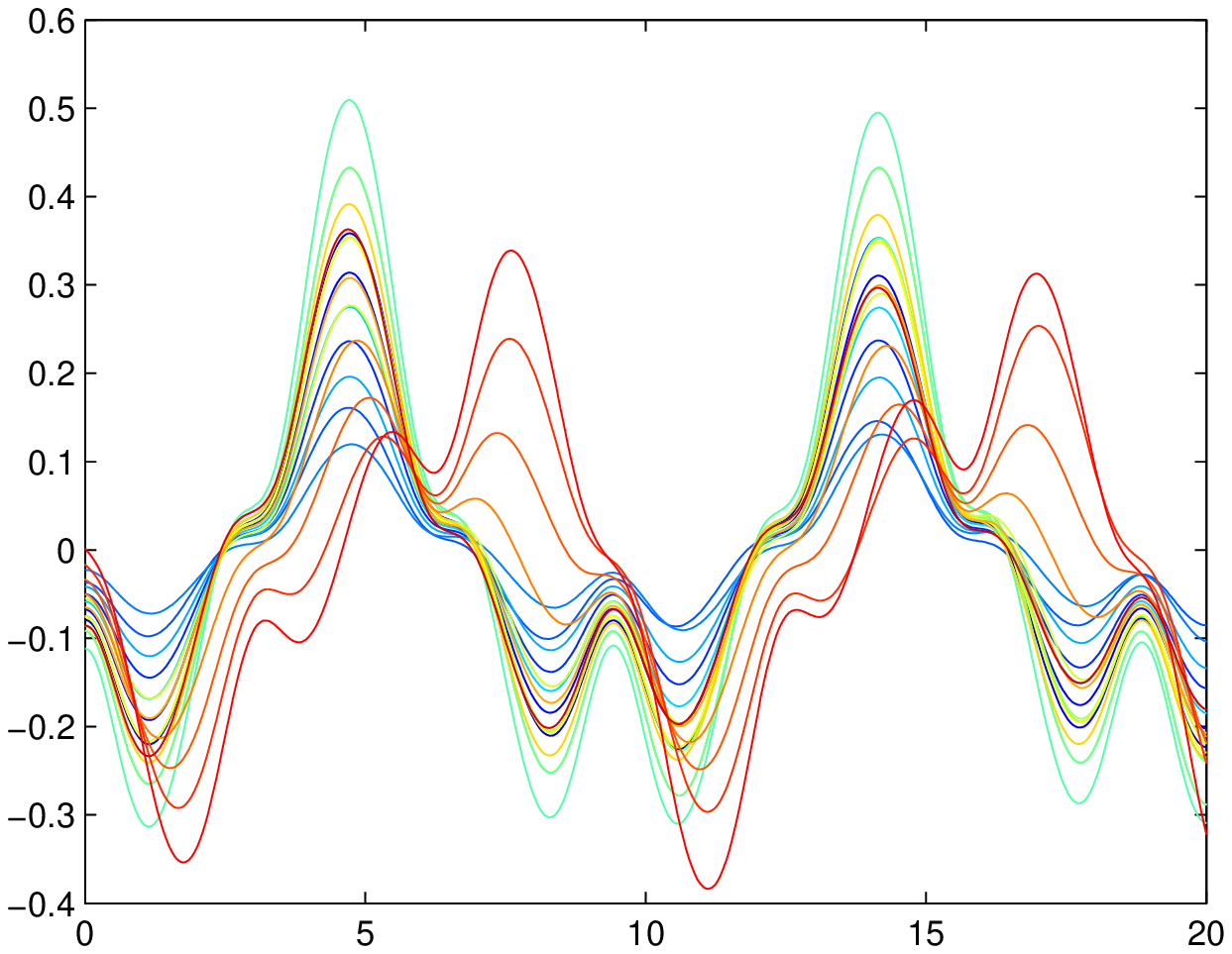}\\
    \includegraphics[width=0.95\linewidth, height = 0.3\textheight]{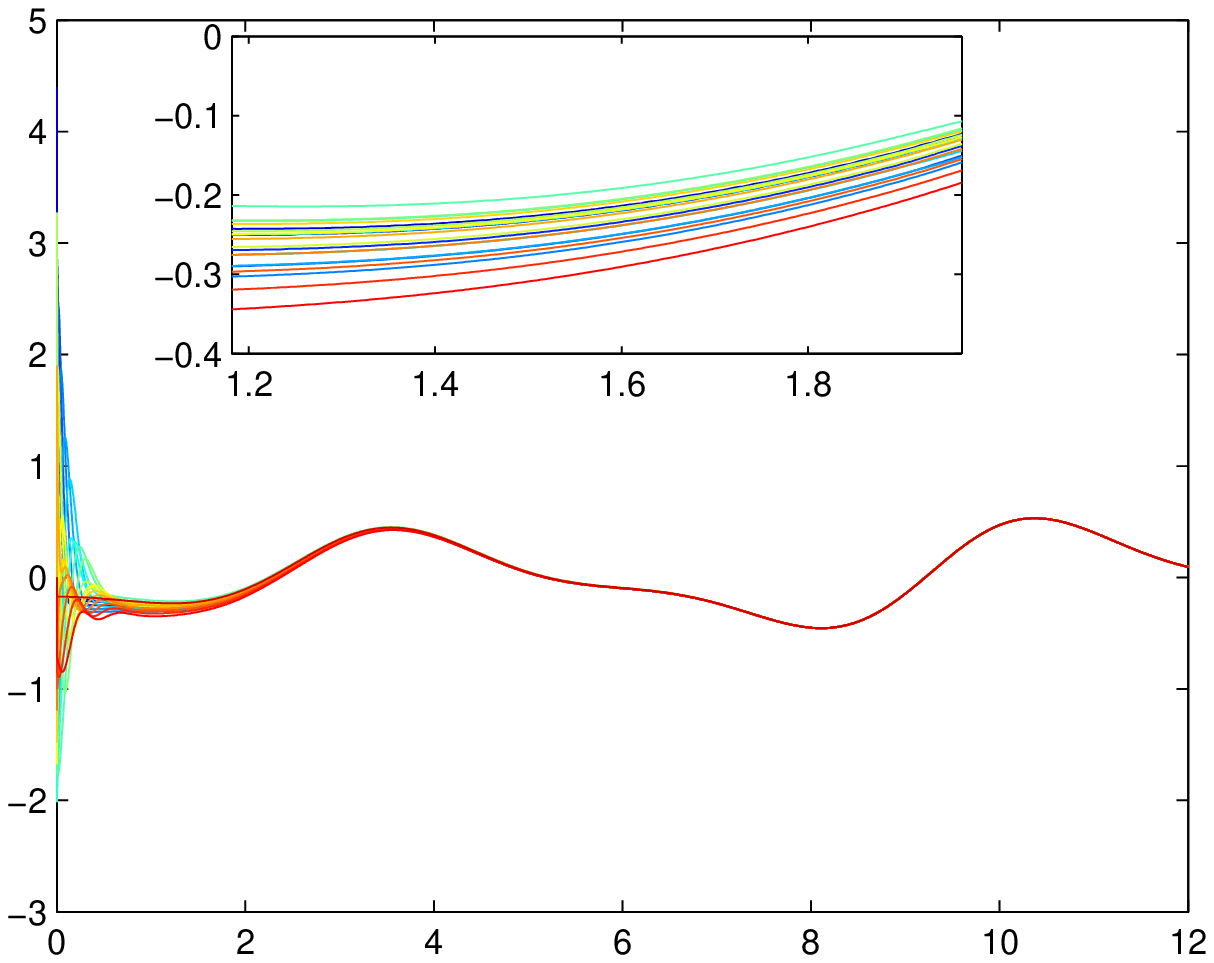} \\
     \includegraphics[width=0.95\linewidth, height = 0.3\textheight]{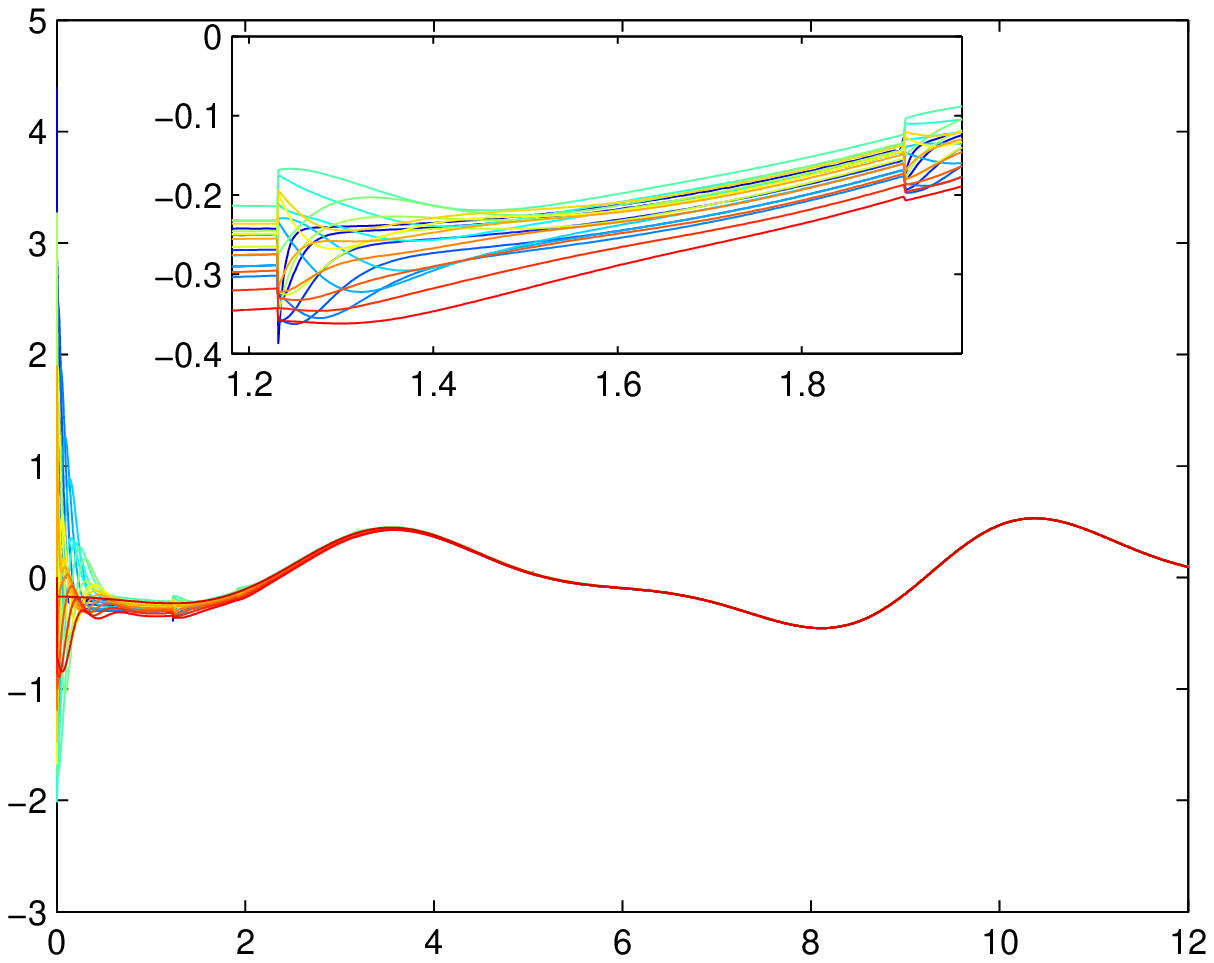}
\caption{Accelerations of the leader and the followers without control
  forces (the top figure), with continuous control forces and switching control forces (the bottom figure).}
   \label{acce}
\end{figure}

Moreover, switching control protocol which does not involve the interpolation technique was simulated in the example to compare difference between the gain scheduling and switching design. The computed performance cost (\ref{cost function}) for the switching control of
the system is equal to $\mathcal{J}(u)=40.2858$. The simulation results
are shown in Fig.~\ref{trackingerror},~\ref{trajectory},~\ref{distance},~\ref{control_signals} and \ref{acce}.

The Fig.~\ref{trackingerror} demonstrates that the proposed continuous protocol
enables all the followers to synchronize to the leader. The displacement and velocity trajectories of each block are shown in Fig.~\ref{trajectory}. Fig.~\ref{distance} shows relative distances between the blocks
$\alpha_i-\alpha_{i-1}+h$, where $h=1$m is the headway distance. From the
figure, the relative distances vary initially between
$0.9$m and $1.1$m and remain positive. This indicates that collisions
between the blocks were avoided. Fig.~\ref{control_signals} confirms that the
control signals are continuous for each follower by using the interpolation
technique, while the switching control scheme leads to discontinuous control
signals. This also causes accelerations of the subsystems to be continuous
and discontinuous, respectively; see Fig.~\ref{acce}. Discontinuous
accelerations are often undesirable in practical systems.
We also notice that the accelerations exerted by the proposed controllers are of
comparable value with the accelerations occurring in the control-free system
(Fig.~\ref{acce}); this indicates that the controller gains have been tuned
to acceptable values.

\section{Conclusions}
\label{conclusions}
The paper has considered the leader-follower control problem for a
parameter varying system with directed communication topology and linear
uncertain coupling, subject to norm-bounded constraints. In contrast to
many existing works, we assume that the leader is selected among the agents, and
remains coupled to the rest of the system. Therefore, in a sense the
problem under consideration has addressed synchronization in leaderless
interconnected multi-agent systems. The main technical challenge
has been to overcome the effects of physical interconnections.

For this problem, we have proposed a continuous gain-scheduled
consensus-type control protocol which employs an interpolation
technique. The sufficient condition for the synchronization of the system
is obtained which guarantees a suboptimal bound on tracking performance.
The condition is in the LMI form, and is numerically tractable, although it
requires solving coupled LMIs.

\section{Acknowledgements}
This work was supported by the Australian Research Council under Discovery Projects funding scheme (projects DP0987369 and DP120102152).


\begin{thebibliography}{99}


\bibitem{Shamma2007}
J. S. Shamma (editor). Cooperative control of distributed multi-agent systems. Wiley, 2007.

\bibitem{[Tuna2008]}
S. E. Tuna. Synchronizing linear systems via partial-state coupling. {\em Automatica} 2008; \textbf{44}(8):2179--2184.

\bibitem{[Tuna2009]}
S. E. Tuna. Conditions for synchronizability in arrays of coupled linear systems. {\em IEEE Transactions on Automatic Control} 2009; \textbf{54}(10):2416--2420.

\bibitem{Olfati2004}
R. Olfati-Saber and R. M. Murray. Consensus problems in networks of agents
with switching topology and time-delays. {\em IEEE Transactions on Automatic Control} 2004; \textbf{49}(9):1520--1533.

\bibitem{Olfati2007}
R. Olfati-Saber, J. A. Fax, and R. M. Murray. Consensus and cooperation in
networked multi-agent systems. {\em Proceedings of the IEEE} 2007; \textbf{95}(1):215--233.

\bibitem{Wieland2011}
P. Wieland, R. Sepulchre, and  F. Allg\"{o}wer. An internal model principle is necessary and sufficient for linear output synchronization. {\em Automatica} 2011; \textbf{47}(5):1068--1074.


\bibitem{Pecora1990}
L. M. Pecora and T. L. Carroll. Synchronization in chaotic systems. {\em Physical Review Letters} 1990; \textbf{64}(8):821--824.

\bibitem{Grip2012}
H. F. Grip, T. Yang, A. Saberi, and A. A. Stoorvogel. Output
synchronization for heterogeneous networks of non-introspective
agents. {\em Automatica} 2012; \textbf{48}(10):2444--2453.


 \bibitem{Siljak1978}
D. D. \v{S}iljak. Large-scale dynamic systems: Stability and structure. NY: North-Holland, New York, 1978.

 \bibitem{Pota2006}
H. R. Pota, G. X. Athanasius, V. Ugrinovskii, and L. Li. Control design for interconnected power systems with OLTCs via
robust decentralized control. \emph{American Control Conference}, Minneapolis, Minnesota, USA, 2006; 3469--3474.

 \bibitem{Siljak1991}
D. D. \v{S}iljak. Decentralized control of complex systems. MA: Academic Press, Boston, 1991.

\bibitem{Shamma2012}
J. S. Shamma. An overview of LPV systems. \emph{Control of Linear Parameter Varying Systems with Applications}, J. Mohammadpour and C. Scherer (editors),  Springer, 2012.


\bibitem{Bianchi2006}
F. D. Bianchi, H. De Battista, and R. J. Mantz. Wind turbine control systems: principles, modelling and gain scheduling design. Springer, 2006.


\bibitem{MSK-2012}
K. Morrisse, G. Solimini, and U. Khan. Distributed control schemes for
wind-farm power regulation. \emph{North American Power Symposium (NAPS)}, Champaign, IL, USA, 2012; 1--6.

\bibitem{Seyboth2012a}
G. S. Seyboth, G. S. Schmidt, and F. Allg\"{o}wer. Cooperative control of linear parameter-varying systems. \emph{American Control
Conference}, Montreal, CA, USA, 2012; 2407--2412.


\bibitem{Liu2013}
Q. Liu, C. Hoffmann, and H. Werner. Distributed control of parameter-varying spatially interconnected
systems using parameter-dependent Lyapunov functions. \emph{American Control Conference}, Washington, DC, USA, 2013; 3278--3283.

\bibitem{[Cheng2013b]}
Y. Cheng, V. Ugrinovskii, and G. Wen. Guaranteed cost tracking for uncertain
coupled multi-agent systems using consensus over a directed
graph. \emph{Australian Control Conference}, Perth, Australia, 2013; 375--378.
Also, see arXiv:1309.0365.


\bibitem{Seyboth2012}
G. S. Seyboth, G. S. Schmidt, and F. Allg\"{o}wer. Output synchronization of linear parameter-varying systems via dynamic couplings. {\em IEEE Conference Decision and Control}, Maui, Hawaii, USA, 2012; 5128--5133.


\bibitem{U8}
V. Ugrinovskii. Gain-scheduled synchronization of parameter varying systems via relative $H_\infty$ consensus with application to synchronization of uncertain bilinear systems. {\em Automatica} 2014; \textbf{50}(11):2880--2887.


\bibitem{U7b-journal}
V. Ugrinovskii. Conditions for detectability in distributed
consensus-based observer networks. {\em IEEE Tran. Autom. Contr.} 2013;
\textbf{58}:2659--2664.


\bibitem{[Yoon2007]}
 M. Yoon, V. Ugrinovskii, and M. Pszczel. Gain-scheduling of minimax optimal state-feedback controllers for uncertain linear parametervarying systems. {\em IEEE Transactions on Automatic Control} 2007; \textbf{52}(2):311--317.

 \bibitem{Scardovia2009}
L. Scardovi and R. Sepulchre. Synchronization in networks of identical
linear systems. {\em Automatica} 2009; \textbf{45}:2557--2562.

\bibitem{Jadbabaie2003}
A. Jadbabaie, J. Lin, and S. A. Morse. Coordination of groups of mobile autonomous agents using
nearest neighbor rules. {\em IEEE Transactions on Automatic Control} 2003; \textbf{48}(6):988--1001.


\bibitem{Ren2007}
W. Ren and E. Atkins. Distributed multi-vehicle coordinated control via local
information exchange. {\em Int. J. Robust Nonlinear Control} 2007; \textbf{17}:1002--1033.

\bibitem{Li2010}
Z. Li, Z. Duan, G. Chen, and L. Huang. Consensus of multiagent systems
and synchronization of complex networks: a unified viewpoint. {\em IEEE
Trans. Circuits Syst. I, Reg. Papers} 2010; \textbf{57}(1):213-"1¤7224.


\bibitem{Hong2006}
Y. Hong, J. Hu, and L. Gao. Tracking control for multi-agent consensus with an active leader and variable topology. {\em Automatica} 2006; \textbf{42}(7):1177--1182.


\bibitem{Hu2007}
J. Hu and Y. Hong. Leader-following coordination of multi-agent systems with coupling time delays. {\em Physical A} 2007; \textbf{374}(2):853--863.


\bibitem{Zhang2012}
H. W. Zhang, Frank L. Lewis, and Z. H. Qu. Lyapunov, adaptive, and optimal design techniques for cooperative systems on directed communication graphs. {\em IEEE Transactions on Industrial Electronics} 2012; \textbf{59}(7):3026--3041.


\bibitem{Stilwell1999}
D. J. Stilwell and W. J. Rugh. Interpolation of observer state feedback controllers for gain scheduling. {\em IEEE Transactions on Automatic Control} 1999; \textbf{44}(6):1225--1229.


\end{thebibliography}
\end{document}